\documentclass[
twocolumn,
tighten,
]{aastex62}

\makeatletter
\renewenvironment{thebibliography}[1]{\global\bibtrue
\ifrnaas\newpage\fi
\onecolumngrid
\vspace{20pt}
\goodbreak
    \hbox to\textwidth{\hss\normalsize REFERENCES\hss}
\vspace{6pt}\parskip=0pt
\twocolumngrid
\par
 \raggedright
\small
\ifmodern\else
 \vspace{10pt plus 3pt}\fi
\par
\topsep=0pt
 \list{}%
   {
     \parindent=0pt \parskip=1pt \parsep=0pt 
     \bibindent=0pt                          %
     \footnotesize
\ifmodern\vskip-12pt
\baselineskip=13pt plus 1pt
\else
\ifdoublespace
\baselineskip=20pt
\else
\baselineskip=10pt plus .01pt \fi\fi \interlinepenalty \@M  
     \frenchspacing    
     \hyphenpenalty=10000
     \itemindent=-1.0em                      %
     \itemsep=0pt                            %
     \listparindent=0pt                      %
     \settowidth\labelwidth{0pt} %
     \labelsep=0pt                           %
     \leftmargin=1.0em
     \advance\leftmargin\labelsep
      \let\p@enumiv\@empty
      }%
    \sloppy\clubpenalty10000\widowpenalty10000%
    \sfcode`\.\@m\relax
}
  {\def\@noitemerr
    {\@latex@warning{Empty `thebibliography' environment}}%
\endlist
    \onecolumngrid 
\global\bibfalse
\newpage
  }
\makeatother

\usepackage{hyperref}

\hypersetup{
	pdftitle={Tidally induced bars in gas-rich dwarf galaxies orbiting the Milky Way},
	pdfauthor={Gajda et al.},
     pdfnewwindow=true,      
     colorlinks=true,    
     linkcolor=xlinkcolor,     
     citecolor=xlinkcolor,     
     filecolor=xlinkcolor,  
     urlcolor=xlinkcolor,      
final=true
}

\usepackage{amsmath}

\shorttitle{Bars in gas-rich dwarfs around the Milky Way}
\shortauthors{Gajda et al.}

\submitjournal{ApJ}
\received{26 June 2018}
\revised{16 October 2018}
\accepted{20 October 2018}

\begin{document}

\title{Tidally induced bars in gas-rich dwarf galaxies orbiting the Milky Way
}

\author[0000-0002-9290-0473]{Grzegorz Gajda}
\affiliation{Nicolaus Copernicus Astronomical Center, Polish Academy of Sciences, Bartycka 18, 00-716 Warsaw, Poland}
\affiliation{Aix Marseille Univ, CNRS, CNES, LAM, Marseille, France}

\author[0000-0001-7138-8899]{Ewa L. {\L}okas}
\affiliation{Nicolaus Copernicus Astronomical Center, Polish Academy of Sciences, Bartycka 18, 00-716 Warsaw, Poland}

\author[0000-0001-6079-1332]{E. Athanassoula}
\affiliation{Aix Marseille Univ, CNRS, CNES, LAM, Marseille, France}

\begin{abstract}
Bars in galaxies may form not only through instability, but also as a result of an interaction with another galaxy. In
particular, they may appear in disky dwarf galaxies interacting with Milky Way-like galaxies. Here we report the results
of $N$-body/SPH simulations of such dwarfs orbiting in the static potential of a larger galaxy. We used several models of
the dwarf galaxy, all of the same mass, but covering a large range of gas fractions: $0$, $30$ and $70\%$. We also
tested the impact of subgrid star formation processes. In all cases bars of similar length formed in the stellar disk of
the dwarfs at the first pericenter passage. However, unexpectedly, the gaseous component remained approximately
axisymmetric and unaffected by the bar potential. The bar properties did not change significantly between two
consecutive pericenters. The impact of the later encounters with the host depends strongly on the exact orientation of
the bar at the pericenter. When the bar is spun up by the tidal force torque, it is also shortened. Conversely, if it
is slowed down, it gets longer. In the models with a low gas fraction the bars were more pronounced and survived until
the end of the simulations, while in the dwarfs with a high gas fraction the bars were destroyed after the second or
third pericenter passage. In terms of the ratio of the corotation radius to the bar length, the bars are slow, and
remain so independently of the encounters with the host.
\end{abstract}


\section{Introduction}

Bars are among the most prominent features of disk galaxies and have strong impact on their evolution and
dynamics. In the Local Universe, about $25-30\%$ of disks host a strong bar \citep[e.g.\ ][]{sheth08, masters11,
cheung13}, but if one includes weak bars, the bar fraction may be as high as $60\%$. This fraction may depend on the
environment and some studies found that it is larger in denser regions \citep{skibba12, mendez-abreu12}. The bar fraction
decreases for galaxies with lower masses \citep{lee12, erwin18}. In particular, \citet{janz12} found that dwarfs in the
Virgo cluster exhibit a bar fraction of $18\%$.

It was realized early on that disk galaxies are often unstable and susceptible to the formation of a bar in their
centers \citep{miller70, hohl71, ostriker_peebles73, miller_smith79}. Soon after their formation, bars undergo buckling
instability \citep{combes_sanders81, combes90, raha91}, which leads to the thickening of their central part and the
formation of a boxy/peanut (b/p) bulge, as reviewed by \citet{athanassoula16}. \citet{erwin_debattista17} found that
more than $50\%$ of barred galaxies exhibit b/p bulges and this fraction is higher for high-mass galaxies. Further
evolution, called secular, is governed by the emission of the angular momentum by the bar and its absorption by the outer
parts of the disk and the dark matter halo. It leads to gradual lengthening and slowing down of the bar
\citep{athanassoula02, athanassoula03}. For more details, the reader is referred to the reviews by
\citet{athanassoula13review} and \citet{sellwood14}.

\citet{contopoulos80} and \citet{athanassoula80} established that the upper limit for the bar length is set by the
corotation radius $R_\mathrm{cr}$, where the bar pattern speed $\Omega_\mathrm{p}$ is equal to the circular frequency.
Later, \citet{athanassoula92a,athanassoula92b} suggested that the ratio of the corotation radius to the bar
length $\mathcal{R}=R_\mathrm{cr}/l_\mathrm{bar}$ should be in a range $1<\mathcal{R}<1.4$. Such bars are called
\emph{fast}, while \emph{slow} bars are defined by $\mathcal{R}>1.4$. Fast bars constitute about two-thirds of bars
with measured pattern speed and almost all of them have $\mathcal{R}<2$ \citep{corsini11, font17}.

Interactions between galaxies are known to influence their structures \citep{toomre_toomre72}. This applies also to
galaxies in groups \citep{mayer01, kazantzidis11, villalobos12, semczuk18} and clusters \citep{mastropietro05, lokas16,
semczuk17}. In particular, bars may form in response to the interaction \citep{noguchi87, gerin90}. Bars may by induced
by a smaller companion \citep[e.g.\ ][]{lang14, pettitt_wadsley18}, a larger one, such as a host galaxy \citep{lokas14,
gajda17} or a cluster \citep{lokas16}, or a galaxy of similar size \citep{lokas18}.

Tidally induced bars and those formed by instability share many properties \citep{berentzen04, gajda16}, but there
are also some differences. If the bar is initially stable against bar formation, in the secular phase its pattern speed
and length remain constant rather than, respectively, decrease and increase \citep{salo91, miwa_noguchi98, gajda17}.
Moreover, the tidally induced bars are usually slow, having $\mathcal{R}\approx 2-3$ \citep{miwa_noguchi98,
berentzen04, aguerri_gonzalez-garcia09, lokas14, lokas16, gajda17, lokas18}. \citet{miwa_noguchi98} argued that this is
due to the angular momentum transfer from the bar region to the perturber. Nevertheless, bars formed spontaneously can
also be slow, provided that their host galaxy is dominated by the dark matter halo \citep{debattista_sellwood00,
lokas16, pettitt_wadsley18}.

Concerning the effect of the gas, initially only the response of gas to the bar potential was studied, mostly in
two-dimensional potentials \citep[e.g.\ ][]{sanders_huntley76, athanassoula92b, kim12, sormani15}. Normally, gas should
follow periodic orbits in a given potential, since gravity dominates over pressure. As gas is the collisional component,
which must have a well-defined velocity at any given point, such orbits cannot intersect. However, in the case of
strong bars the orbits of the x$_1$ family may have cusps or loops at their ends \citep{athanassoula92a}. Additionally,
nested orbits of this family can also intersect. As a result, offset shocks form on the leading sides of the bars, which
correspond to dust lanes in observed barred spiral galaxies \citep{athanassoula92b}. Thus a gas parcel spends more time
there than on the trailing side. The gas then loses angular momentum and inflows toward the center of the galaxy,
where it forms a circumnuclear disk, in which gas follows x$_2$ orbits. Due to the inflow, a region with some gas
deficiency is created in less than one full bar rotation \citep{kim12}.

The gas flowing towards the center of the galaxy can form a central mass concentration (CMC) which may lead to the bar
dissolution \citep{pfenniger_norman90}. This indeed can happen for bars, as was found by early simulations
\citep{berentzen98, bournaud_combes02, berentzen04}. A small gas fraction has negligible impact on the bar evolution
\citep{villa_vargas10}, but if it is larger, then the bar evolution is different. In gas-rich galaxies bars are formed
later and are weaker than in gas-poor ones \citep{shlosman_noguchi93, berentzen07, villa_vargas10, athanassoula13gas}.
The buckling instability does not take place, but instead the bar swells symmetrically \citep{debattista06,
berentzen07} right after the bar formation. In the secular phase, bars in gas-rich galaxies do not grow and keep their
pattern speed constant.

Cosmological simulations are the only way of taking into account fully the effect of the environment. Such work was first performed by \citet{kraljic12} and \citet{scannapieco_athanassoula12}. \citet{kraljic12} studied a suite of zoom-in simulations and found trends consistent with the ones observed by \citet{sheth08}. The bars started to appear around $z\approx 1$ and their fraction increased with decreasing redshift. However, it is often difficult to disentangle if the bars were triggered by instability or by tidal interactions \citep{spinoso17, zana18, peschken_lokas18}. Large samples of galaxies can be only followed in large cosmological boxes, however their resolution might be not sufficient to resolve properly the inner structure \citep{algorry17, peschken_lokas18}.

Dwarf galaxies are smaller, but more numerous than large galaxies. They exhibit certain differences when compared to
the latter, for example in their scaling relations \citep{mateo98, tolstoy09, mcconnachie12}. Dwarfs can be divided in
roughly two groups: dwarf spheroidals (dSph) and dwarf irregulars (dIrr). The dIrr galaxies are gas-rich and exhibit
certain amount of rotational support, whereas the dSphs are gas-poor and are supported by random motions of stars. In
the Local Group of galaxies the population of dwarfs exhibits a density-morphology relation. Close to the Milky Way and
Andromeda one finds predominantly dSphs, while dIrrs are located mostly far away from the big spirals. Such a relation
points to a possible evolutionary link between the two groups.

In the tidal stirring scenario for the formation of dSphs, initially disky dwarfs are captured by larger hosts (like
the Milky Way) and transformed into spheroids by the tidal forces \citep{mayer01, kazantzidis11, lokas11}. A bar forms
in the disk of the dwarf galaxy during the first encounter with the host \citep{klimentowski09}. If this scenario is
valid, one can expect hints of bar presence in some of the dwarfs in the Local Group. Indeed, the elongated shapes of
Sagittarius, Ursa Minor and Carina dwarf \citep{lokas10, lokas12, fabrizio16} may be remnants of bars. Furthermore,
bar-like shapes can be found in ultra-faint dwarfs Hercules and Ursa Major II \citep{coleman07, munoz10}. Of course,
tidally induced bars may form not only in the Local Group, but also in other dwarf galaxies interacting with their
hosts and in clusters \citep{aguerri_gonzalez-garcia09}.

In this paper we extend the work presented in \citet{gajda17} and earlier in \citet{lokas14}. Compared to these works,
we take into account the impact of the interstellar medium, modelled using the Lagrangian method called Smoothed
Particle Hydrodynamics \citep[SPH,][]{lucy77, gingold_monaghan77, monaghan92}. We investigate the impact of different
fractions of the gaseous component with respect to the total baryonic mass. Rather large gas fractions are used in
order to mimic dIrrs. In some of the runs we also include the processes of cooling, star formation and feedback
modelled with a sub-grid prescription.

In Section~\ref{sec_simulations} we introduce our simulations and describe the construction of initial conditions. The
results of the computations are presented in Section~\ref{sec_results}. First, we describe the general properties of
the dwarf galaxies and the evolution of their bars. Then we focus on the impact of tidal torques on the bar strength
and rotation. Finally, we briefly discuss the star formation rate and the distribution of young stars. In
Section~\ref{sec_discussion} we discuss our results and compare them to the existing literature and then conclude in
Section~\ref{sec_conclusions}.

\section{Simulations}
\label{sec_simulations}

\subsection{The code}

The simulations were performed using \textsc{Gadget3} code, which is a descendant developer version of a well-known,
publicly available \textsc{Gadget2} \citep{springel05}.
We use the code in two different setups, which we call \emph{non-star forming} (noSF) and \emph{star forming} (SF).
In the noSF mode the gas physics is dictated solely by hydro- and thermodynamics.
The equation of state of the ideal gas is used and the gas evolves adiabatically (i.e.\ the gas is \emph{not} isothermal).
In this mode we do not use any subgrid prescription.

In the SF mode we did use subgrid prescriptions for cooling, star formation and supernova feedback.
The implementation of these processes followed the
model of \citet{springel_hernquist03}. It includes radiative cooling of the hydrogen and helium as in \citet{katz96},
down to $10^4$~K.
The star formation prescription, which includes supernova feedback, is based on a hybrid two-phase
model. The gas above a density threshold of $n_\mathrm{H}=0.18$ cm$^{-3}$ (eq.\ 23 of \citealt{springel_hernquist03}) is treated as if it consisted of cold, molecular clouds
and a hot phase, excited by feedback.
Such gas follows the equation of state given by equation (13) of \citet{springel_hernquist03}.
Only gas with density above the threshold is allowed to form stars. We did not include galaxy winds in our simulations, as their parameters were developed for much larger galaxies.
We do not include metal-line cooling in order to keep consistency with the assumed model of star formation and feedback.
We set the parameters of the subgrid physics to the values given in \citet{springel_hernquist03}.

\subsection{Initial conditions}

The simulation setup was similar to the one used by \citet{lokas14} and \citet{gajda17}, however with certain
modifications. We constructed a model of a dwarf galaxy consisting of a spherical Navarro-Frenk-White
\citep[NFW,][]{nfw95} dark matter halo and a baryonic disk. The dark matter halo had a virial mass of $10^9$ M$_{\sun}$
and a concentration of 20. For the baryonic disk of mass $2 \times 10^7$ M$_{\sun}$, we first generated an $N$-body
model of an exponential disk with a radial scale-length of $0.41$ kpc and a vertical scale-height of $0.082$ kpc. To
create the models we used the procedures of \citet{widrow_dubinski05} and \citet{widrow08}.
Both components consisted of $4\times 10^6$ particles.
Such a large number of particles is believed to be sufficient to follow the interaction and bar
evolution reliably \citep{gabbasov06, dubinski09}, as well as the effect of the gas \citep{patsis_athanassoula00}.

The mass we chose for the dwarf galaxy is of the order of the masses of the classical dwarf satellites of the
Milky Way \citep{mcconnachie12} once we take into account the tidal stripping. The dwarfs beyond the virial radii of
bigger galaxies or in isolation are usually gas rich \citep{mcconnachie12, papastergis12, sales15}. According to
\citet{papastergis12} galaxies at our mass range have on average the gas fraction of the total baryonic mass at a level
of $\sim 80\%$. In the Local Group and its vicinity dwarf irregulars have gas fractions ranging from $40\%$ (IC
1613) to $90\%$ (NGC 3109) (\citealt{mcconnachie12}, Fig. 8). We chose two levels of gas fraction for our simulations,
namely $70\%$ as it is the most common value and $30\%$ to understand better the effect of gas fraction growth and
mimic what could happen if some of the gas was ram-pressure stripped.

We used the following procedure to build the initial conditions for the gas-rich dwarfs. In the previously generated
\mbox{$N$-body} disk we randomly changed a given fraction of stellar particles into gaseous particles, setting their
temperature to $2000$ K. For this temperature, the resulting gaseous disk was prone neither to collapsing nor to
thickening when seen in the edge-on view. We also changed the velocities of these new gaseous particles in order not to
induce thickening by the initial velocity dispersion. The radial and vertical velocities were set to zero, while the
angular velocity was set to the circular velocity at a given distance from the galaxy center at the galaxy mid-plane
(i.e.\ we did not vary it with the height).

Finally, we evolved in isolation the configuration consisting of the halo, the stellar disk and the gas disk. To create
initial conditions for runs which did not include the star formation processes, we ran the simulation in isolation for
$3$ Gyr, until most of the initial transients disappeared. To prepare the star forming runs we performed a slightly
different procedure. First, the dwarf was evolved for $1.5$ Gyr \emph{without} star formation. Only then we turned on
the star formation processes and evolved it for the next $1.5$ Gyr. For completeness, we evolved also the purely
stellar case in isolation for $3$~Gyr.

\begin{table}
\centering
\caption{Basic properties of the simulations.
$Q_\mathrm{min}$ was calculated after $3$ Gyr period of settling down in isolation.
}
\label{tab_simulations}
\begin{tabular}{ccccc}
\hline\hline
Run & Gas fraction & Star formation & Line color & $Q_\mathrm{min}$\\
\hline
Nbody   &  $0\%$ & --- & Red    & $4.5$\\
g30noSF & $30\%$ & No  & Green  & $3.0$\\
g30SF   & $30\%$ & Yes & Blue   & $4.3$\\
g70noSF & $70\%$ & No  & Violet & $1.9$\\
g70SF   & $70\%$ & Yes & Brown  & $5.4$\\
\hline
\end{tabular}
\end{table}

We performed five different simulation runs.
One of them was a collisionless $N$-body case, without any gaseous component.
It was intended to serve as a fiducial, comparison model.
For the gas-rich runs we used two levels of gas fraction: $30\%$ and $70\%$ of baryons, as the dwarf galaxies far away from large galaxies are known to possess large amounts of gas \citep{mcconnachie12, papastergis12, sales15}.
For each level of the gas fraction we performed two simulations. In
one of them the star formation processes were included, while in the other one they were turned off.
The properties of the simulation runs are summarized in Table~\ref{tab_simulations},
where we listed gas fractions and whether we employed subgrid recipes for the cooling, star formation and feedback.
The naming scheme reflects these choices.
In the fourth column of the table we show the line colors which is used consistently throughout this paper
to illustrate the results for the different runs in all the figures.

\begin{figure}
\centering
\includegraphics{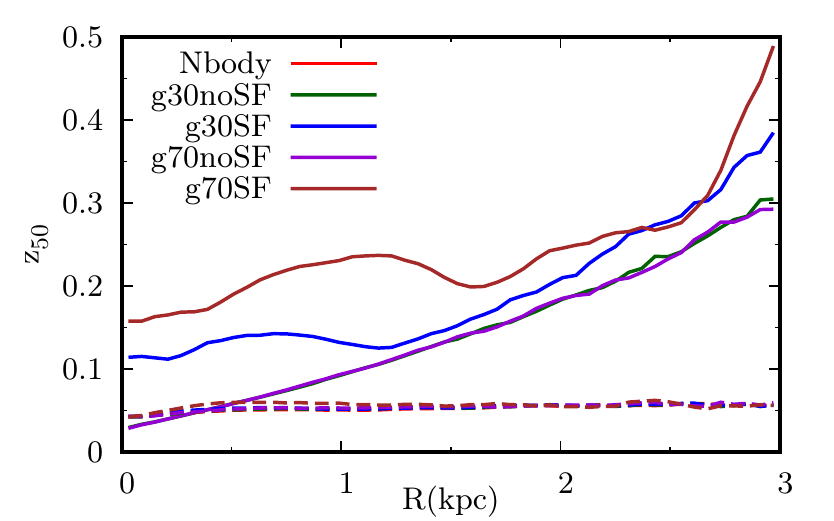}
\caption{Radial profiles of the $z_{50}$ parameter in the initial conditions, separately for gas (solid lines) and stars
(dashed lines).}
\label{fig_initial_thickness}
\end{figure}

We evaluated axisymmetric stability of the models using the criterion given by equation (9) of \citet{romeo_wiegert11}.
The calculations were done after the initial evolution period for all dwarfs. 
The minimum values of the $Q$ parameter are listed in the last column of Table \ref{tab_simulations}.
According to \citet{romeo_wiegert11} the axisymmetric stability criterion as usual reads $Q>1$.
In order to check if the models are stable in isolation against bar formation, we evolved the models of the dwarf
galaxies in isolation for an additional, much longer time. The models corresponding to the runs Nbody, g30noSF, and
g70noSF were evolved for $7$ Gyr, while those for the runs g30SF and g70SF were run for $10$~Gyr. This way we checked
that the models were indeed stable against spontaneous bar formation for such time periods.

\begin{figure}
\centering
\includegraphics{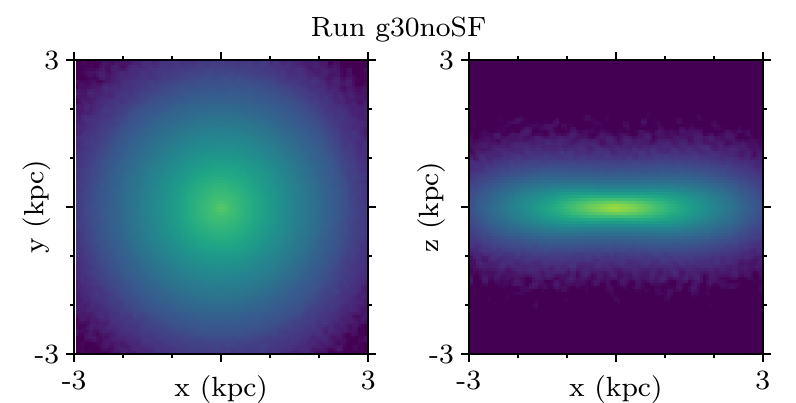}
\includegraphics{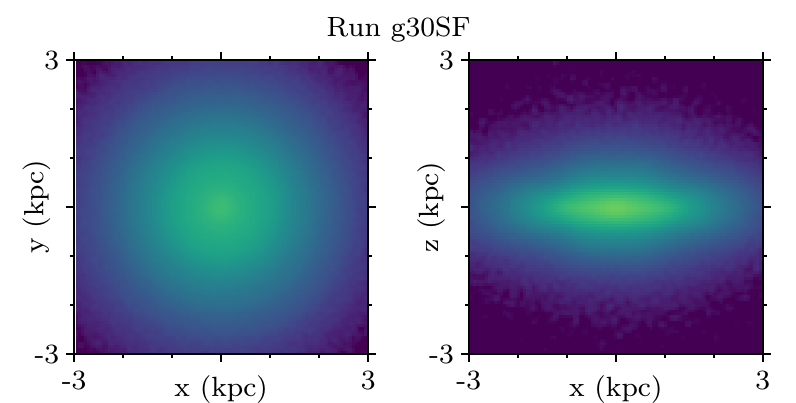}
\includegraphics{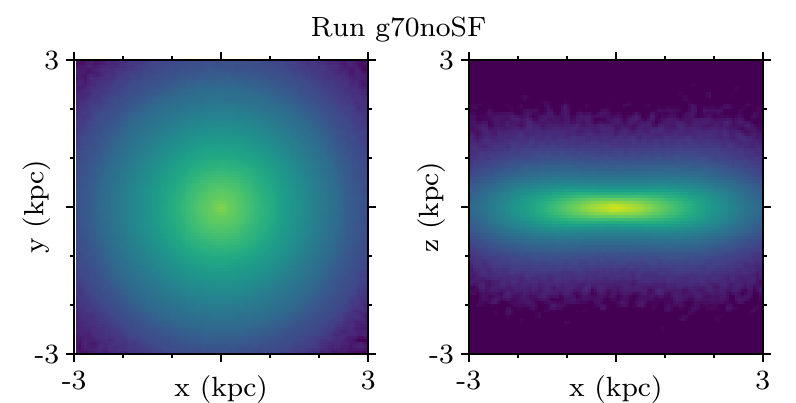}
\includegraphics{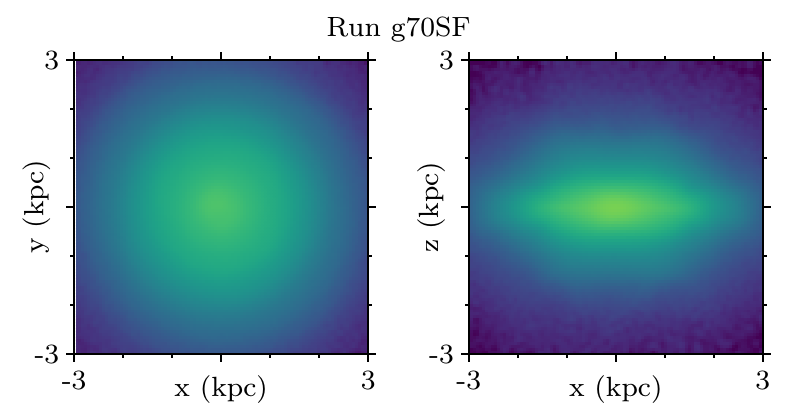}
\caption{
Face-on (left column) and edge-on (right column) views of surface density at the beginning of the simulations.
Each row corresponds to one of the runs with gas included.
}
\label{fig_initial_map}
\end{figure}

To give an idea on the initial distribution of the gas, in Figure \ref{fig_initial_thickness} we plot $z_{50}$, the
median value of the vertical coordinate $|z|$ of particles at a given radius. The stars exhibited a flat distribution
with radius in all of our initial conditions. On the other hand, the distribution of the gas was flaring with
increasing radius. Note that this is an expected behavior found also in other studies, see for example
\citet{benitez-llambay18}. The star forming initial conditions were also initially thicker than the non-star forming
ones, which is due to the continuous input of the feedback energy.

In Figure~\ref{fig_initial_map} we also show gas surface densities after the initial evolution in isolation.
In the face-on view one can notice that the gaseous disks are axisymmetric.
In the edge-on view we see what was already illustrated by Figure~\ref{fig_initial_thickness}: the gaseous disks in the star-forming runs are thicker than in the non-star forming
ones and the thickest one is present in model g70SF.

\subsection{Setup of the runs}

These dwarf galaxy models were then evolved around a Milky Way-like host. The host was modelled as a~spherically
symmetric static potential of an NFW halo with $7.7\times 10^{11}$ M$_{\sun}$ mass and concentration of $27$. All the
dwarfs were initialized at a~$120$~kpc apocenter of an elongated orbit, with a pericenter at $24$~kpc. One may be
worried about the lack of dynamical friction \citep{chandrasekhar43}, which could change the orbits of the dwarfs.
However, at such a mass ratio between the host and the satellite the orbital decay is small \citep[e.g.\ ][]{frings17},
as confirmed by the earlier works \citep{lokas14, gajda17}. A much more important omission was the lack of a hot gas
halo of the host galaxy, which resulted in the lack of ram-pressure stripping of the gas from the dwarf. We decided not
to include this effect in order to focus on the bar evolution under tidal interactions alone. The disks of the dwarfs
were placed in the orbital plane, in such a way that the orbital motion was prograde. Such a~configuration is known to
maximize the effect of the interaction \citep[e.g.\ ][]{toomre_toomre72, lokas15, gajda17, lokas18}.

The evolution of the dwarf galaxies was followed for $10$ Gyr and outputs were saved every $0.05$ Gyr. The softening
for dark matter particles was $0.03$ kpc, while for stellar and gaseous particles it was equal to $0.01$ kpc.

\section{Results}
\label{sec_results}
\subsection{General properties}

For each snapshot of the simulations we computed the center of mass of the dwarf galaxy. We used the common algorithm
based on successively shrinking spheres \citep{power03}. Next, we determined the principal axes of the stellar
component diagonalizing iteratively the shape tensor computed using particles inside an ellipsoid of a $1$ kpc long
major axis \citep[see][]{zemp11, gajda15}. For most measurements we aligned the reference frame with the principal axes
of the dwarfs' stellar component.

\begin{figure}
	\centering
	\includegraphics{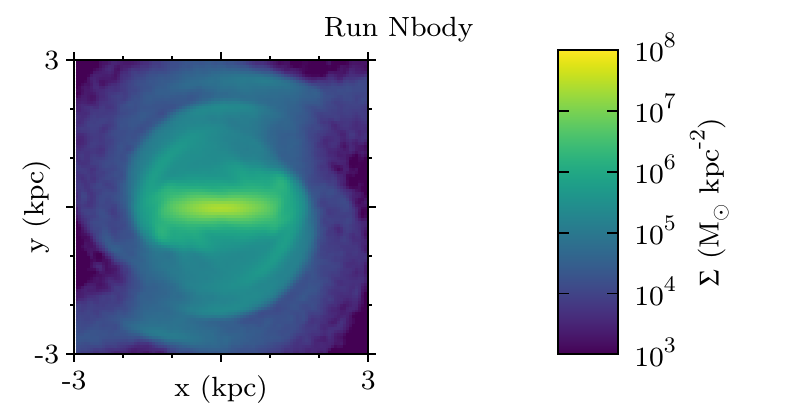}
	\includegraphics{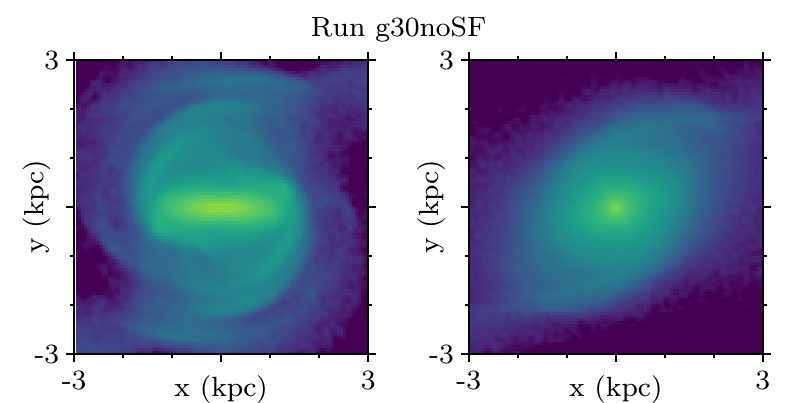}
	\includegraphics{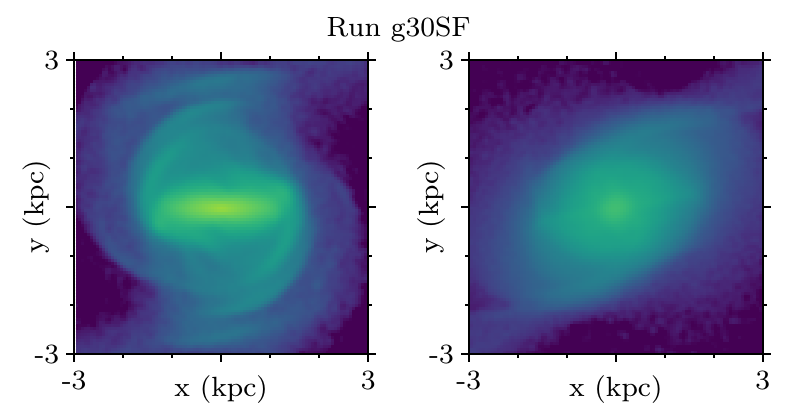}
	\includegraphics{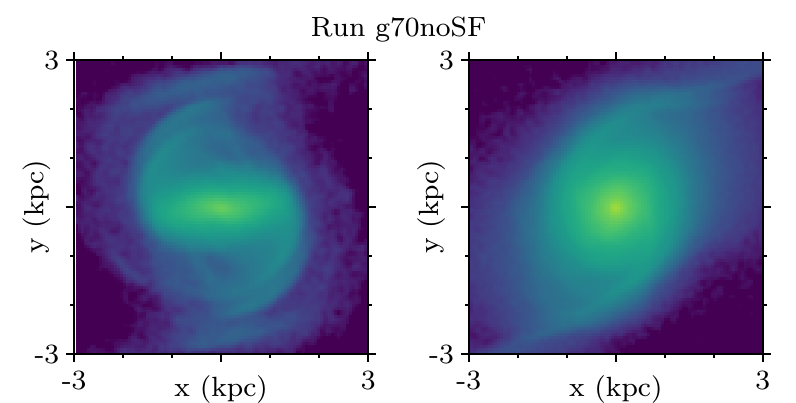}
	\includegraphics{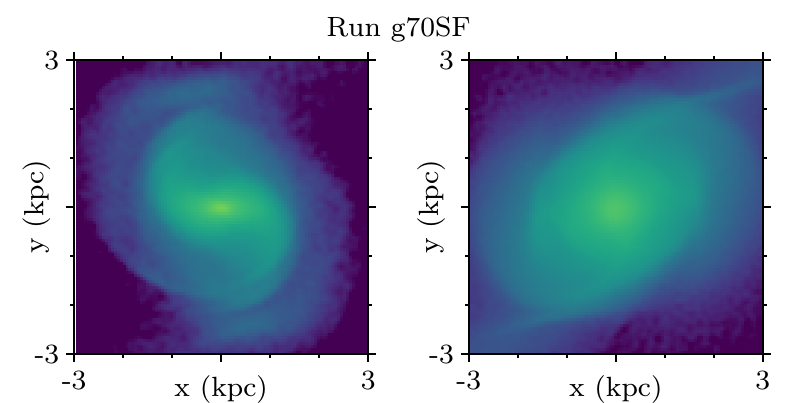}
	\caption{
	Stellar (left column) and gas (right column) surface densities at the time of the second apocenter ($t=2.2$~Gyr).
	Each row corresponds to one of the runs.
	}
	\label{fig_density_map}
\end{figure}

In Figure~\ref{fig_density_map} we show the surface density maps for the stellar and gaseous components at the time of
the second apocenter ($2.2$ Gyr). When comparing the plots one should remember that the initial surface densities of
both components varied between the runs, as the total baryonic mass was assigned in different ways to stars and gas. In
all the simulations bars formed in the stellar component and they all had similar lengths. However, there was no trace
of the bar or its impact in the gaseous component. With an increasing gas fraction the stellar bars became less
elongated and more round. In the star forming runs the central gas surface densities were lower than in the pure
hydrodynamical runs. This is due to the stars forming from the gas where its density exceeds the threshold value. In
fact, in the non-star forming simulations a peak in the gas density was formed, presumably due to the driving of the
gas toward the center by the bar.

\begin{figure}
\centering
\includegraphics{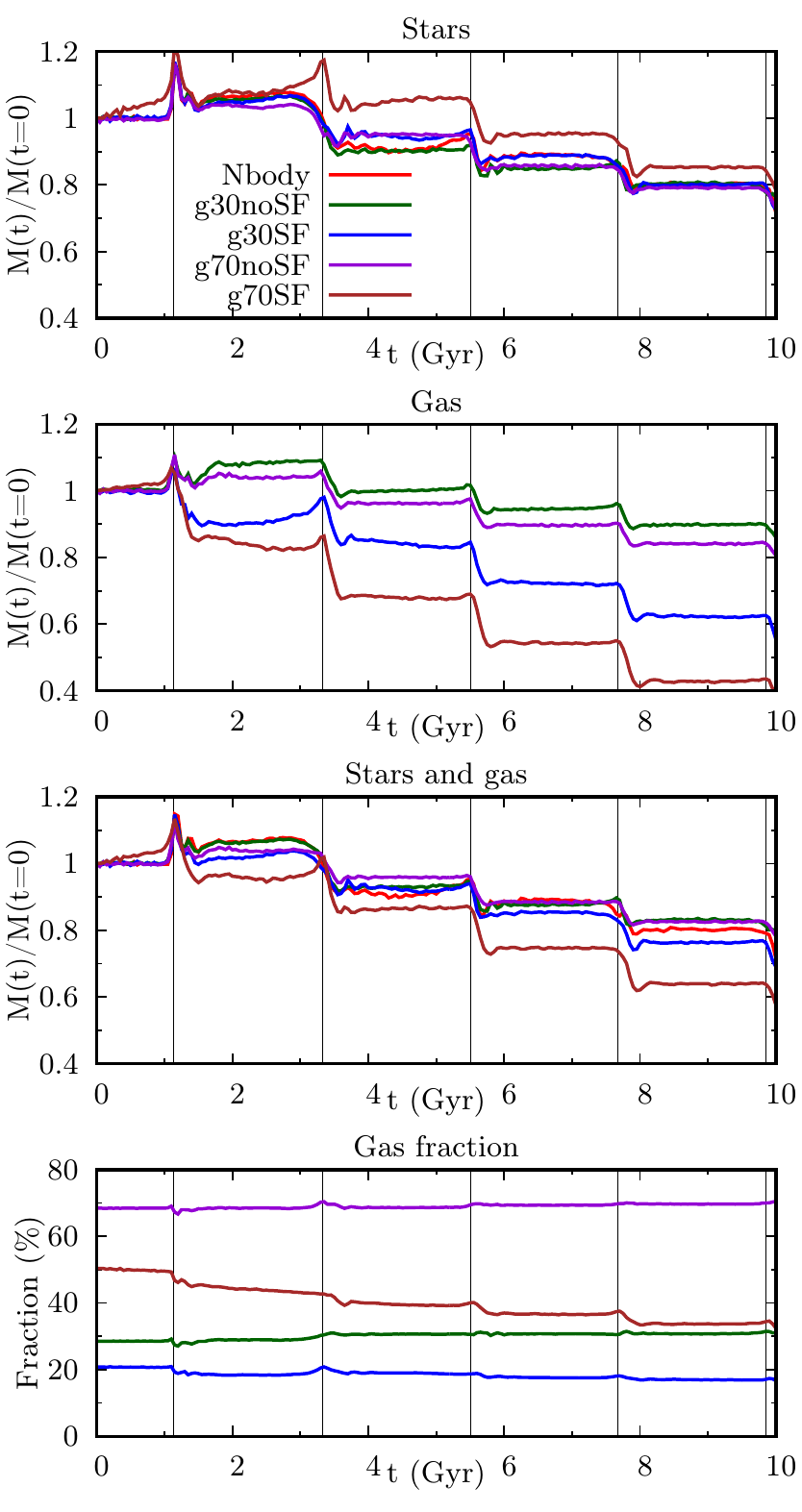}
\caption{Evolution of the baryonic content inside a sphere of radius $0.5$ kpc, normalized by the initial value. The
first panel shows the stellar content, the second one the gas, the third one the total baryonic content and the last
one the fraction of the baryonic mass in gas. Vertical lines indicate pericenter passages of the dwarfs.}
\label{fig_mass}
\end{figure}

In Figure \ref{fig_mass} we show how the baryonic content in the central $0.5$ kpc of the dwarfs evolved. At the first
pericenter the stellar mass usually rose slightly due to the rearrangement of matter into a bar. Later on, the stellar
mass dropped at the pericenters due to tidal stripping and remained constant between the encounters with the host. The
gas mass in the non-star forming runs also slightly grew at the first pericenter. When the stars were allowed to form,
the gas content decreased. During further evolution, the behavior of the gas was similar to the one of stars: the mass
was constant between the pericenters and dropped during the encounters with the host.

\begin{figure*}
\centering
\includegraphics{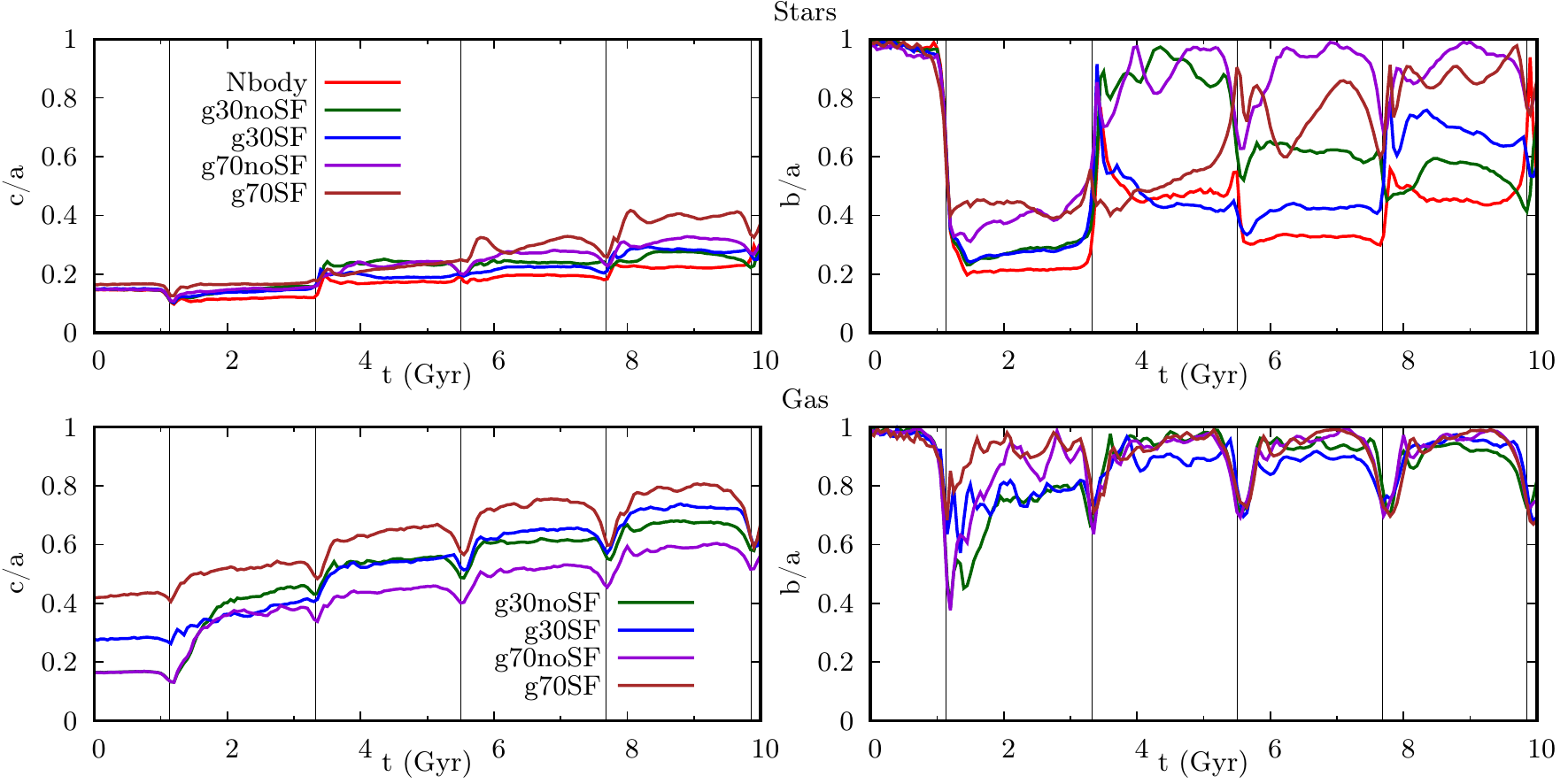}
\caption{Evolution of the axis ratios of the stellar (top row) and gaseous (bottom row) component. In the left column
we plot the thickness ($c/a$) and in the right one the elongation ($b/a$). The shapes have been calculated at the radius
of $1$ kpc. The vertical lines mark the times of the pericenter passages.}
\label{fig_shape}
\end{figure*}

It is also interesting to look at the relative behavior of the two components, as expressed by the gas fraction. The
first thing one notices is that initially the gas fractions of the star-forming runs were lower than in their analogues
that did not form stars. This was of course due to the transformation of the gas into stars which took place during the
evolution in isolation. When measured over the whole body of the dwarfs, the initial gas fractions were much closer to
the assumed ones, as the star formation happened mostly in the centers of the dwarfs.
During further evolution, the gas fraction in g30SF and g70SF dropped further.
The steady decrease between the first and the second pericenter passage was due to the on-going creation of stellar
particles. However, later on the gas fraction drops only at the pericenter passages, when there is hardly any star formation taking
place (see Fig.~\ref{fig_sfr}).
We thus we ascribe the drop to the enhanced stripping of the gas in the star forming models. We
believe that this is a result of a more extended distribution of the gas in these models, as can be seen in Fig.
\ref{fig_initial_map}.

In the two non-star forming runs the gas fraction slightly rose at the pericenters and the final difference was of the order of a few per cent. It seems that the gaseous component was more tightly bound and therefore harder to strip than the stellar material.
The main reason for the difference in the gas fraction in the central $0.5$~kpc between the star forming and
non-star forming runs is the continuous transformation of gas into stars in the former.
However, more efficient stripping of gas in the star forming runs has also a non-negligible effect, as can be deduced by comparing the behavior at the pericenters.

In Figure \ref{fig_shape} we track the evolution of the shape of the dwarf galaxies. For this purpose, we calculated
the minor-to-major ($c/a$) and intermediate-to-major ($b/a$) axis ratios at the scale of $1$~kpc. The first can be
identified with the thickness of the given component, while the second with its elongation.
Initially, the stellar disks were thin and thickened only a little during subsequent evolution, mainly at the pericenters.
The thickening of the stellar disc depends on the gas fraction in such a way that the stellar discs of dwarfs
with higher gas content thickened more strongly.
The gas disks without star formation (g30noSF and g70noSF) were also initially thin, while the ones with star formation included (g30SF and g70SF) were thicker.
The $c/a$ ratio grew mostly at the pericenters, but also between them. The star forming disks remained thicker
than the no-SF ones, but in the pure hydrodynamical runs the disks puffed up considerably at the first pericenter.

All the dwarfs started as axisymmetric disks, having $b/a\approx 1$. At the first pericenter passage they were
significantly stretched, indicating a bar formation. The stellar components in all the simulations remained elongated,
but more so at lower gas fractions. However, the gaseous components returned to approximate axisymmetry, as we have
seen in the surface density maps (Figure \ref{fig_density_map}). The gas components in the runs with a lower gas
fraction ($30\%$) were slightly more elongated, but after the second pericenter passage all of them were rather round,
with $b/a>0.8$. The evolution of $b/a$ in the stellar component after the second pericenter passage was much more
complicated. Apparently, there were no visible trends. The elongation in the Nbody run changed at each pericenter, it
grew and declined subsequently. The star forming runs became almost round after the second pericenter, but later on in
g30SF the elongation grew considerably. At the end of the simulations, the dwarfs in the runs with high ($70\%$) gas
fraction seemed to be rounder than in the runs with less gas. We will explain the reason for this behavior in
Subsection \ref{subsec_torques}.

\subsection{Evolution of the bars}

\begin{figure}
\centering
\includegraphics{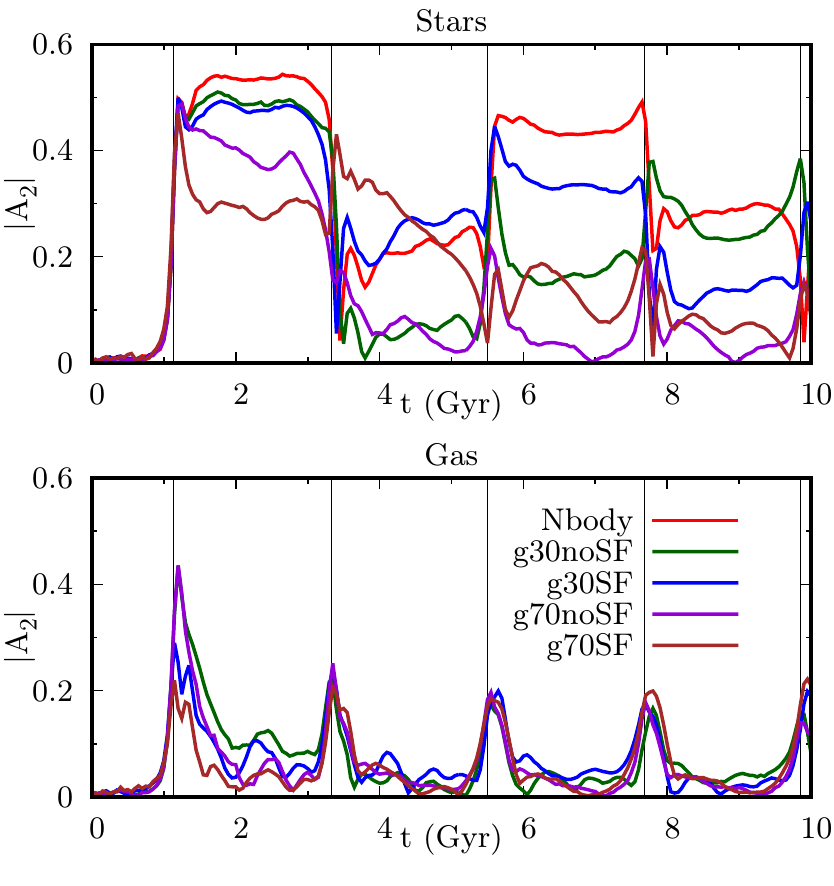}
\caption{The average $|A_2|$ mode measured inside $1.5$ kpc cylinder for stars (top) and gas (bottom). Vertical lines
mark the pericenter passages.}
\label{fig_a2_tot}
\end{figure}

In order to describe the evolution of the bars, we followed the bar mode, defined as
\begin{equation}
A_2=\frac{1}{N}\sum\limits_{j=1}^{N}\exp(i\, 2\varphi_j),
\end{equation}
where we sum over all $N$ particles in the region of interest, $\varphi_j$ are their position angles and $i$ is the
imaginary unit. We calculated the absolute value $|A_2|$ inside the cylindrical radius of $R=1.5$ kpc. We chose this
specific value as it encompasses almost the whole extent of the bars after the first pericenter. We did not use the
more common parameter $A_{2,\mathrm{max}}$ (i.e.\ the maximum of $|A_2(R)|$), as due to the formation of tidal tails the
profiles of the bar mode do not always have a pronounced and well defined maximum \citep[see e.g.\ ][]{lokas14, gajda17}.
We also decided against calculating the bar parameters inside the bar length, which we also estimate later on, because
its evolution could make the interpretation of the results less clear.

The evolution of $|A_2|$ is depicted in Figure \ref{fig_a2_tot}, for both the stellar and the gaseous component. The
behavior of the bar mode reflected the evolution of the $b/a$ ratio.
Assuming the commonly used threshold of $|A_2|>0.2$ for the bar existence, we can see that in all cases
the bars in the stellar component formed at the
first pericenter.
The strongest bar was induced in the Nbody run, and their strength diminished with a growing gas
fraction. The gaseous component was stretched during the encounter with the host galaxy, but it rearranged itself
afterwards so that no sign of any bisymmetric perturbation was present at later times.

The evolution of the bars after the second pericenter was complicated and diverse. In the Nbody run the bar mode was
constant between the pericenter passages, but changed abruptly during the encounters. Similarly, for the runs with a
low gas fraction ($30\%$) $|A_2|$ did not evolve significantly during the time between the pericenter
passages while varying considerably at the times of the encounters with the host. Interestingly, in the high gas
fraction runs the bar mode decreased significantly between the encounters with the host. In g70noSF the decrease took
place between the first and the second pericenter, while in g70SF the bar strengthened at the second pericenter and
then was destroyed almost completely before the third one.

\begin{figure}
\centering
\includegraphics{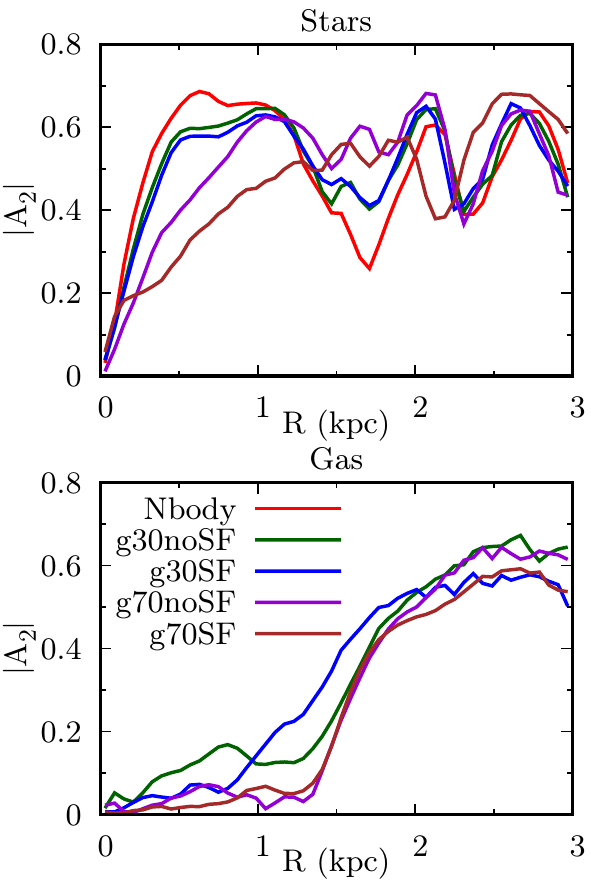}
\caption{Radial profiles of $|A_2|$ for stars (top) and gas (bottom) at the second apocenter ($t=2.2$ Gyr). }
\label{fig_a2_profiles}
\end{figure}

In order to study the structure of the bars we calculated the bar mode in cylindrical shells. Profiles of $|A_2(R)|$ at
the time of the second apocenter are shown in Figure \ref{fig_a2_profiles}. In the stellar component the bar mode
increases with increasing radius in the center, reaches a~maximum value and then drops a little. In the bars formed in
isolation the drop is usually much stronger \citep[e.g.\ ][]{athanassoula_misiriotis02}. The peaks at larger radii are
related to the presence of two sets of spiral arms or escaping shells of matter, which can be seen in Figure
\ref{fig_density_map}. The slope of the inner part of the $|A_2|$ profiles increases with the decreasing gas fraction,
which means that the inner parts of the gas-rich dwarfs were rounder, presumably due to the increased central
concentration of the gas-rich models. In the star forming runs the slope was smaller than in their counterparts without
star formation, probably due to the formation of new stars or different initial vertical distribution of the gas.

As discussed previously, the gas distribution remained near-axisymmetric in the central regions of the dwarfs. For the
gas, $|A_2|$ starts to grow approximately at $R=1.5$ kpc, which corresponds roughly to the end of the bar. In the runs
with a lower gas fraction, the gas was slightly elongated in the center. This might be related to the higher
gravitational impact of the stellar component. The gas distribution at the outskirts was elongated, but it was not
related to the bars, but rather to the tidal disturbance and transition to the tidal tails. Such behavior in the center
of a galaxy is different than what was usually reported in the previous work \citep[e.g.\ ][]{athanassoula92b,
athanassoula13gas} and we discuss this issue later on.

\begin{figure*}
\centering
\includegraphics{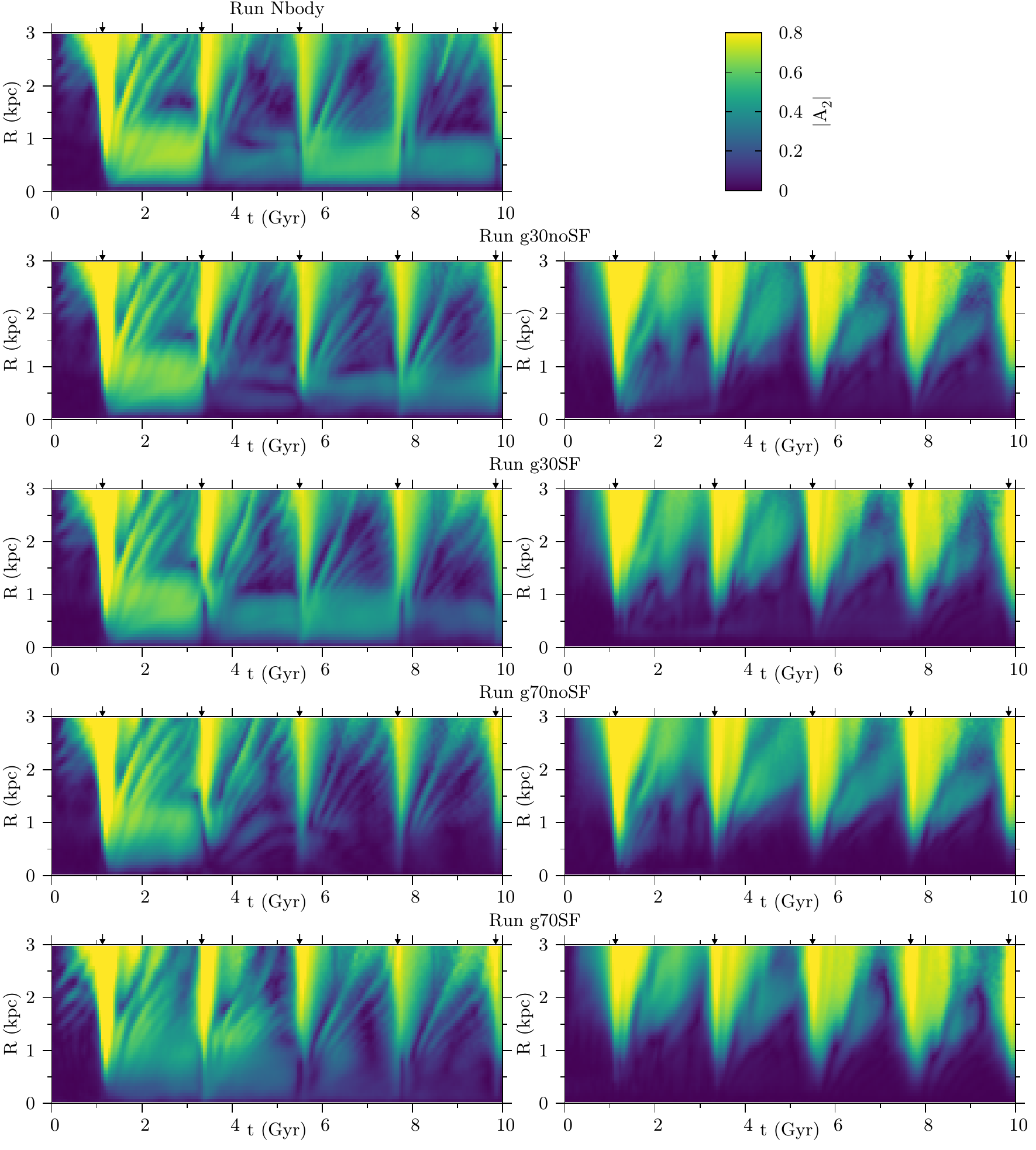}
\caption{Radial profiles of the bar mode $|A_2|$ as a function of time for stars (left) and gas (right).
Each row corresponds to one of the runs.
Small arrows indicate the times of the pericenter passages.
Note that at these periods the color scale is saturated (i.e.\ $|A_2|>0.8$).}
\label{fig_a2_map}
\end{figure*}

The profiles of the bar mode obviously evolved with time, hence in Figure \ref{fig_a2_map} we show values of $|A_2|$ as
a function of time and distance from the center of the dwarf. The most obvious feature of the plots is the elongation
of both the stellar and gaseous component during the pericenter passages, when $|A_2|>0.8$. Between the pericenters one
can notice, in the stellar component, a band of higher values of the bar amplitude at small radii, corresponding to the
bar. However, the gas in the central parts of the dwarfs does not show any trace of bisymmetric perturbation at any
time. Stripes of higher $|A_2|$ travelling outwards correspond to stripped shells of matter.

A large diversity of the evolutionary histories is clearly visible in these plots. In the runs Nbody and g30SF the bars
were weakened at the second pericenter, then they were strengthened at the next one and again weakened at the fourth
one. The bar in the g30noSF model was first weakened and then it grew at the next two pericenters. The life of bars in
the high gas fraction runs was much shorter. In g70noSF the bar was destroyed after the second encounter with the host
galaxy and in g70SF it declined strongly after the second pericenter passage and later was also destroyed.

\begin{figure}
\centering
\includegraphics{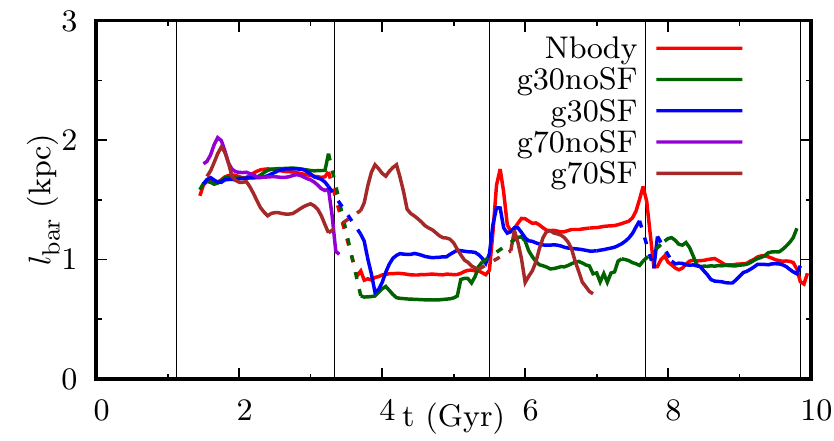}
\caption{The bar length as a function of time. The lines were smoothed over three consecutive outputs. During the
pericenter passages (vertical lines) the measurements were not possible so earlier and later measurements were
connected by dashed lines. All the lines start at the moment when we were able to determine the bar length. The
lines for runs with a high gas fraction end when the amplitude of the bar is too low to measure the length accurately.}
\label{fig_bar_length}
\end{figure}

A feature related to the bar strength is its length. To estimate this property we used the method described previously
in \citet{gajda17}. In short, we looked for the radius where the triaxiality parameter of the stellar component,
measured in elliptical shells, falls below $90\%$ of its maximal value. The evolution of the bar length for all the
runs is presented in Figure \ref{fig_bar_length}. The curves begin at the time of the creation of the bar and for the
high gas fraction dwarfs they end when the amplitude of the bar is too low to measure length accurately. We mark with
dashed lines the periods when the determination of the extent of the bars was not possible due to the elongation of the
whole body of the dwarf. As can be perceived also in Figure \ref{fig_density_map}, initially the bar lengths were
almost the same in all simulations and only the one in run g70SF was slightly shorter. We point out that the values of
$|A_2|$, both integrated and in terms of the profiles, vary from one simulation to another. The bars in the runs with
little or no gas were shortened at the second pericenter passage and their further evolution was correlated with the
evolution of the bar mode in Figure~\ref{fig_a2_tot}. In these runs the bar length seems to remain constant between
pericenter passages. The bars in the runs with high gas fraction declined over time. It is especially interesting for
run g70SF, in which the bar shortened and was destroyed gradually between the second and the third pericenter.

\begin{figure}
\centering
\includegraphics{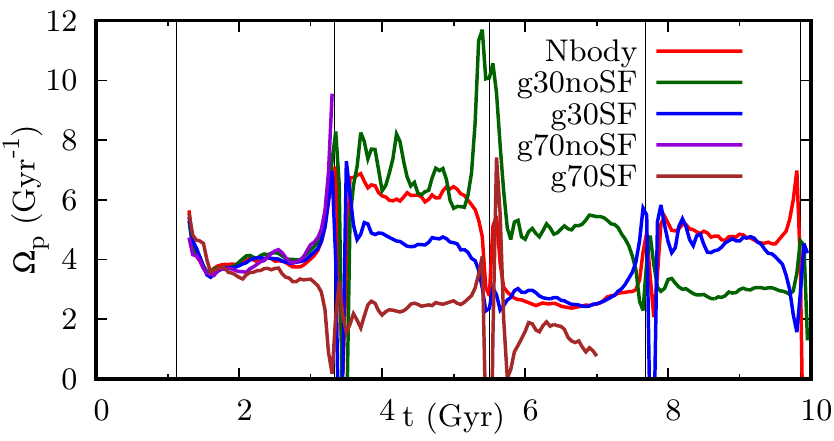}
\caption{The pattern speed of the bars as a function of time. Vertical lines indicate pericenter passages.}
\label{fig_omega_p}
\end{figure}

In Figure \ref{fig_omega_p} we plot the evolution of the pattern speed $\Omega_\mathrm{p}$ of the bars in the dwarfs.
We measured it using the method described in the Appendix of \citet{gajda17}. In short, in order to obtain
$\Omega_\mathrm{p}$ at the time of output $n$ we calculated the $A_2$ phase difference between outputs $n+1$ and $n-1$,
carefully taking into account possible tilting or precession of the disk. Similarly to the bar lengths, the pattern
speeds were almost exactly the same for all runs after the bar formation during the first pericenter passage. Later,
during further encounters with the host, the measured $\Omega_\mathrm{p}$ strongly varied due to the stretching of the
dwarfs induced by the tidal force. The pattern speeds remained rather constant between pericenter passages and evolved
significantly during the encounters with the host galaxy. The changes were anticorrelated with the evolution of the bar
strength and length, as expected \citep{athanassoula03}. The bars were spun up when they were weakened and shortened.
When they were strengthened and elongated, they slowed down. The best example of this behavior is provided by the bar
in the Nbody dwarf, which was speeded up and weakened at the second and fourth pericenter, while during the third
pericenter it was slowed down and strengthened.

\begin{figure}
\centering
\includegraphics{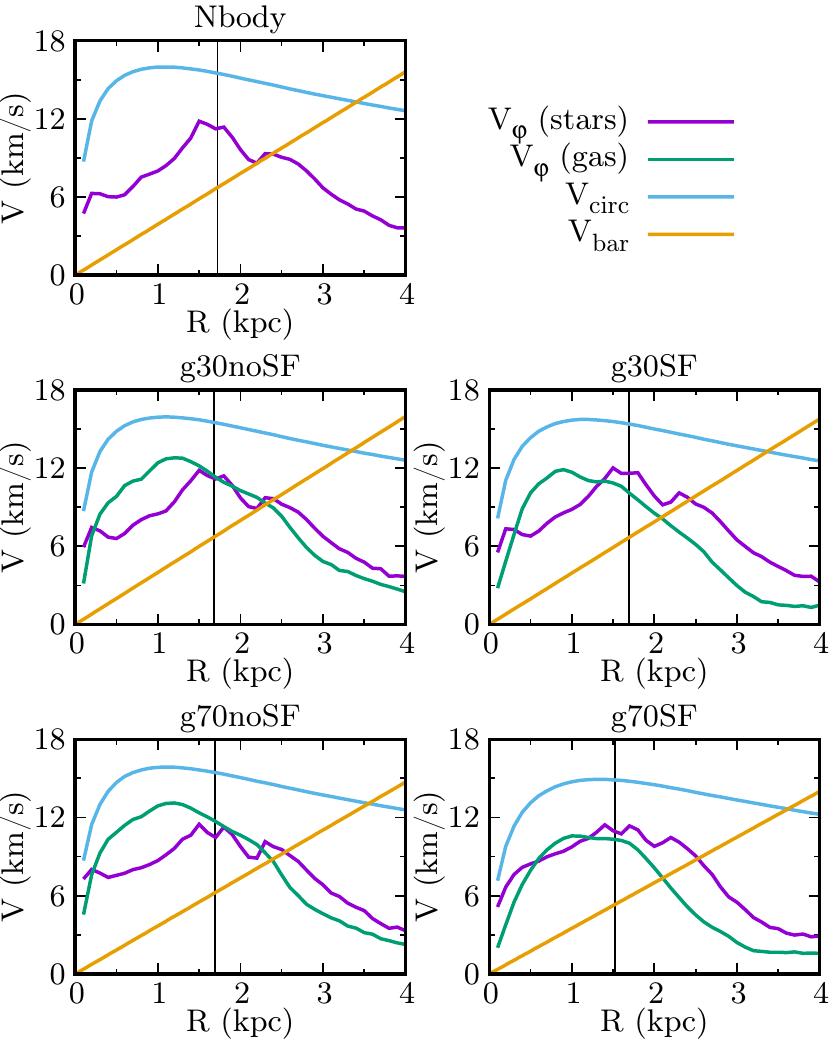}
\caption{Profiles of the rotation velocity of the stars (violet) and the gas (green) together with the circular
velocity (blue) at the second apocenter ($t=2.2$ Gyr). The yellow lines correspond to the linear velocity of the bar
$V_\mathrm{bar}(R)=R\Omega_\mathrm{p}$. Vertical lines indicate the bar lengths.}
\label{fig_cr}
\end{figure}

In Figure \ref{fig_cr} we plot the various velocity curves of the dwarf galaxies at the second apocenter. The rotation
curves $V_{\varphi}$ for the stars and gas were computed by averaging the tangential velocities of all particles
of a~given type at a given cylindrical distance $R$, restricting the measurements to the particles obeying $|z|<1$ kpc.
The circular velocity curve at a cylindrical radius $R$ was computed as
\begin{equation}
V_\mathrm{circ}(R)=\sqrt{\frac{GM(r<R)}{R}},
\end{equation}
where $G$ is the gravitational constant, $R$ is the distance from the center of the dwarf and as $M(r<R)$ we take all
the mass (both dark and baryonic) within a distance $R$ from the center. In addition, we plot the linear velocity of
the bar at a given radius $V_\mathrm{bar}=R\Omega_\mathrm{p}$.
At the beginning of the simulations the rotation velocities of the stars and gas were quite close to the circular
velocity, which had a maximum of $\approx 20$ km s$^{-1}$ at $R\approx 2$~kpc.

After the first pericenter, the circular velocities in all the runs were similar, which means that similar amount of matter was stripped from the dwarfs.
Both baryonic components lost significant amount of rotation, however this happened in a different way for stars and gas.
The decrease of the rotation speed was due to the tidal interaction and creation of the bar, which made the potential non-axisymmetric.
The rotation of the gas in the intermediate part of the bar was faster than for stars simply because the stellar component is kinematically hotter than the gaseous one.
In the outer parts of the dwarfs ($R\approx
2$~kpc) gas was rather elongated (see Figure \ref{fig_a2_profiles}) and beyond $R\approx 3$~kpc the tidal tails were
present. Both of these may be responsible for the fact that the apparent rotation velocity for stars was larger than for
the gas. The overall loss of rotation for the stellar component was smaller for a larger gas fraction.

\begin{figure}
\includegraphics{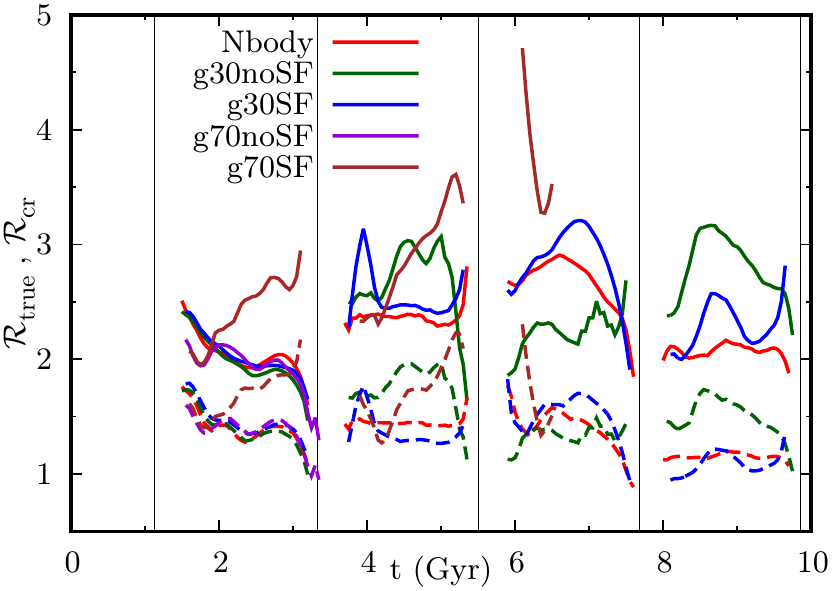}
\caption{Ratio of the corotation radius (solid lines) and the true corotation (i.e.\ radii at which $V_{\varphi\mathrm{,
stars}}=R\Omega_\mathrm{p}$, dashed lines) to the bar length. Measurements were smoothed over three consecutive outputs
and gaps are due to the inability to determine one of the involved quantities. Vertical lines correspond to the
pericenter passages.}
\label{fig_cr_ratio}
\end{figure}

We note that the bar lengths were significantly shorter than the corotation resonance, signifying that our bars are
slow, as in \citet{gajda17}. To track the evolution of the ratio of the corotation radius to the bar length, we
prepared Figure \ref{fig_cr_ratio}. With the solid lines we depicted
$\mathcal{R}_\mathrm{cr}=R_\mathrm{CR}/l_\mathrm{bar}$, i.e.\ the ratio of the corotation resonance (i.e.\ radii at which
$V_\mathrm{circ}=R\Omega_\mathrm{p}$) to the bar length. Dashed lines show the ratio of the true corotation (i.e.\ radii
at which $V_{\varphi\mathrm{, stars}}=R\Omega_\mathrm{p}$) to the bar length, $\mathcal{R}_\mathrm{true}$. One can
notice these curves vary much more between the pericenters than the respective ones for the bar length and the
pattern speed. The reason for this is that $\mathcal{R}$ is a ratio of two numbers that can both vary strongly.

The corotation resonance ratio $\mathcal{R}_\mathrm{cr}$ was always larger than the true corotation ratio
$\mathcal{R}_\mathrm{true}$, because the rotation velocity of the stellar component was smaller than the circular
velocity. We note that $\mathcal{R}_\mathrm{true}$ never dropped below $1$, meaning that on average the stars inside
the bar always moved faster than the bar itself, i.e. there were no large groups of stars on retrograde orbits. Despite
the large changes of the pattern speed, the bar length and the stripping at the pericenters, the ratio
$\mathcal{R}_\mathrm{cr}$ always remained in the range $2$-$3$, meaning that the bars were slow at all times. The
ratios did change after each pericenter. These changes, however, do not seem to be correlated with either spinning up
or slowing down of the bars.

\subsection{Weakening and strengthening of the bars}
\label{subsec_torques}

\begin{figure}
\centering
\includegraphics{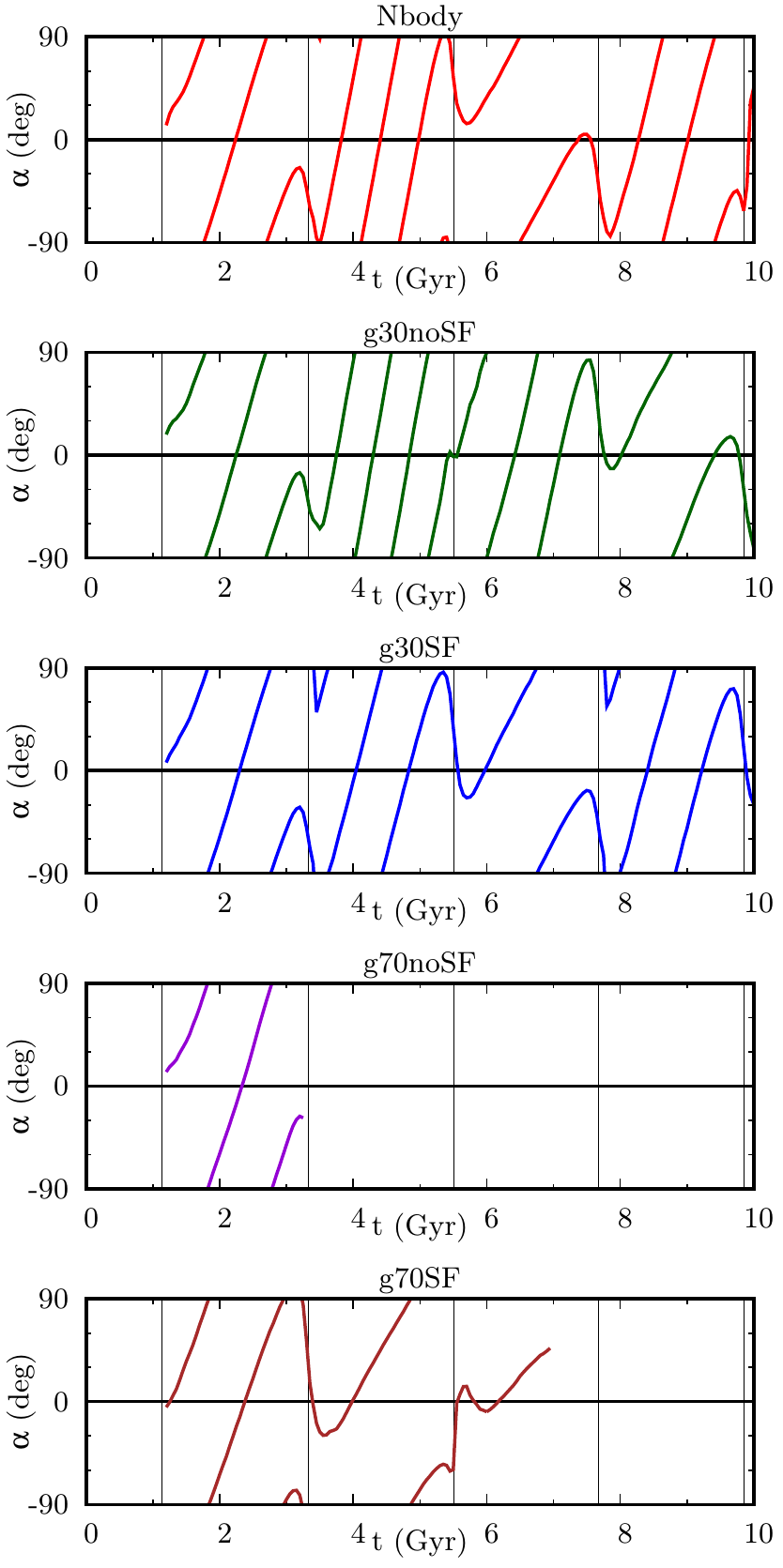}
\caption{The angle $\alpha=\varphi_\mathrm{bar}-\varphi_\mathrm{MW}$ between the bar and the direction to the Milky Way,
as a function of time. Negative angles indicate that the bar is lagging behind the direction to the host, the zero
value that it is aligned with it and the positive mean that it has overtaken the direction to the host. Vertical lines
mark the times of the pericenter passages.}
\label{fig_bar_position_angle}
\end{figure}

\begin{figure}
\centering
\includegraphics{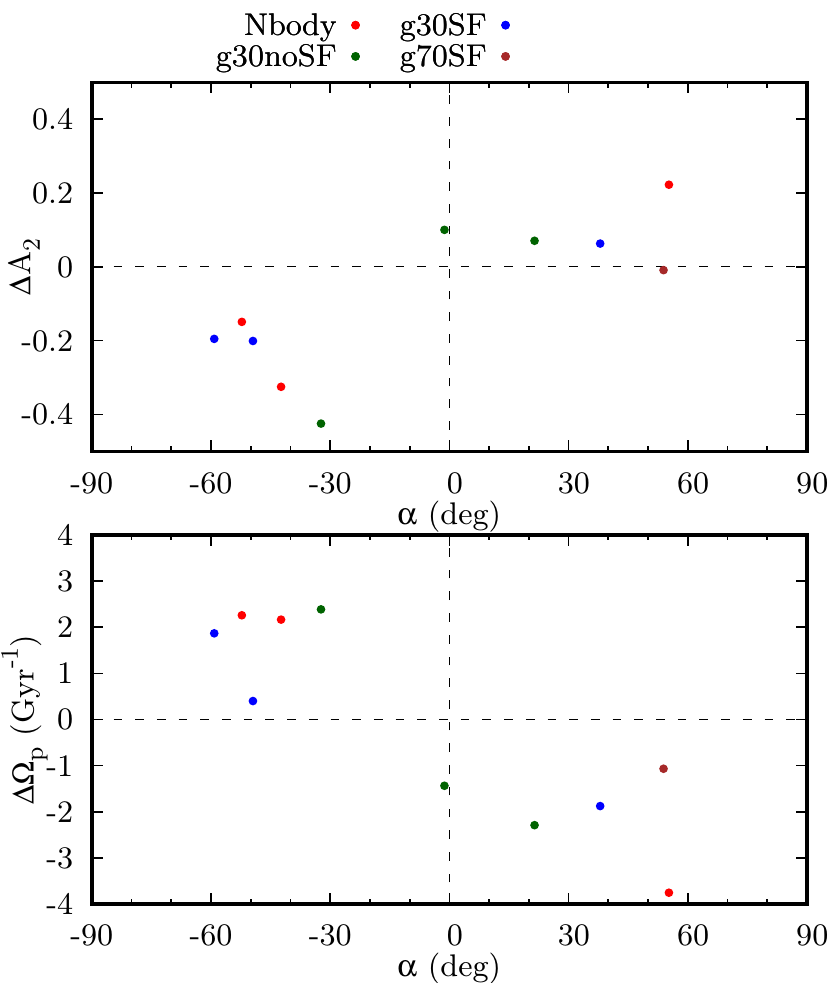}
\caption{
Changes during the pericenter passage of the bar strength (top panel) and the pattern speed (bottom panel) as a function of angle $\alpha=\varphi_\mathrm{bar}-\varphi_\mathrm{MW}$ between the bar and the direction to the Milky Way during that pericenter. Note that g70noSF is not included and for g70SF only one pericenter is.
}
\label{fig_peri_changes}
\end{figure}

\begin{figure*}
\centering
\includegraphics{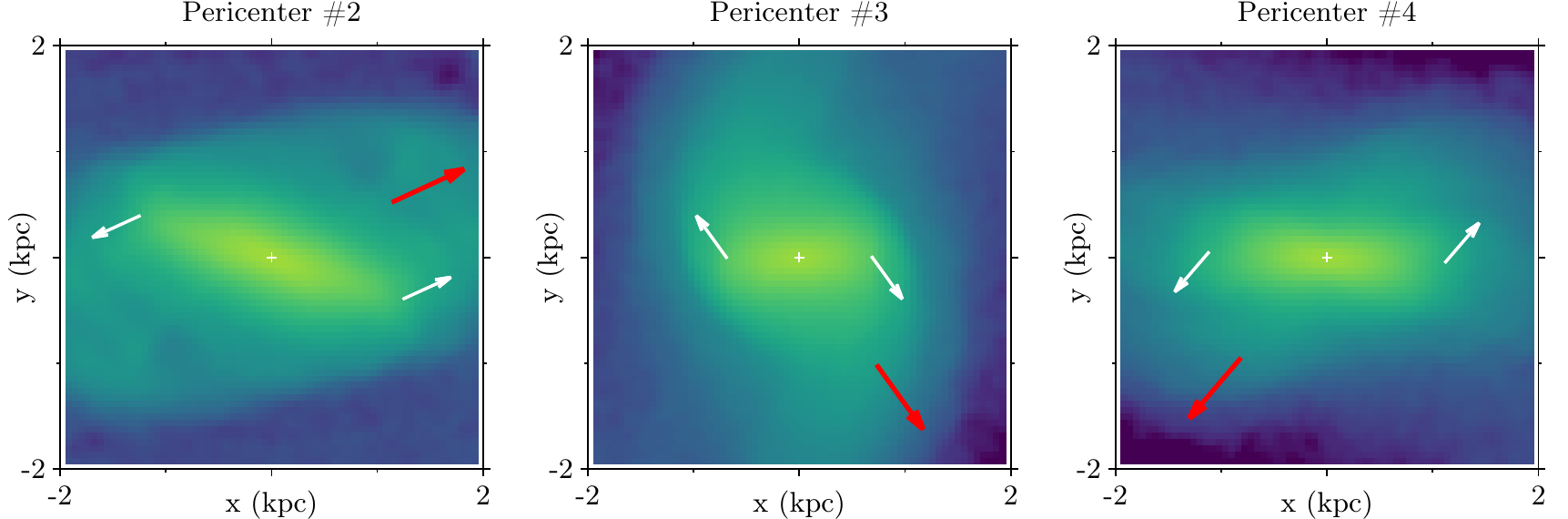}
\caption{Influence of the tidal torque on the bar in the Nbody run. The three panels show the surface density of the
stellar component of the dwarf galaxy during subsequent pericenter passages. The bars rotate in the counter-clockwise
direction. Red arrows mark the direction towards the center of the host galaxy. White arrows indicate approximate
orientation of the tidal force acting on the bar. White crosses mark the centers of the dwarfs.}
\label{fig_tidal_torque}
\end{figure*}

\begin{figure}
\centering
\includegraphics{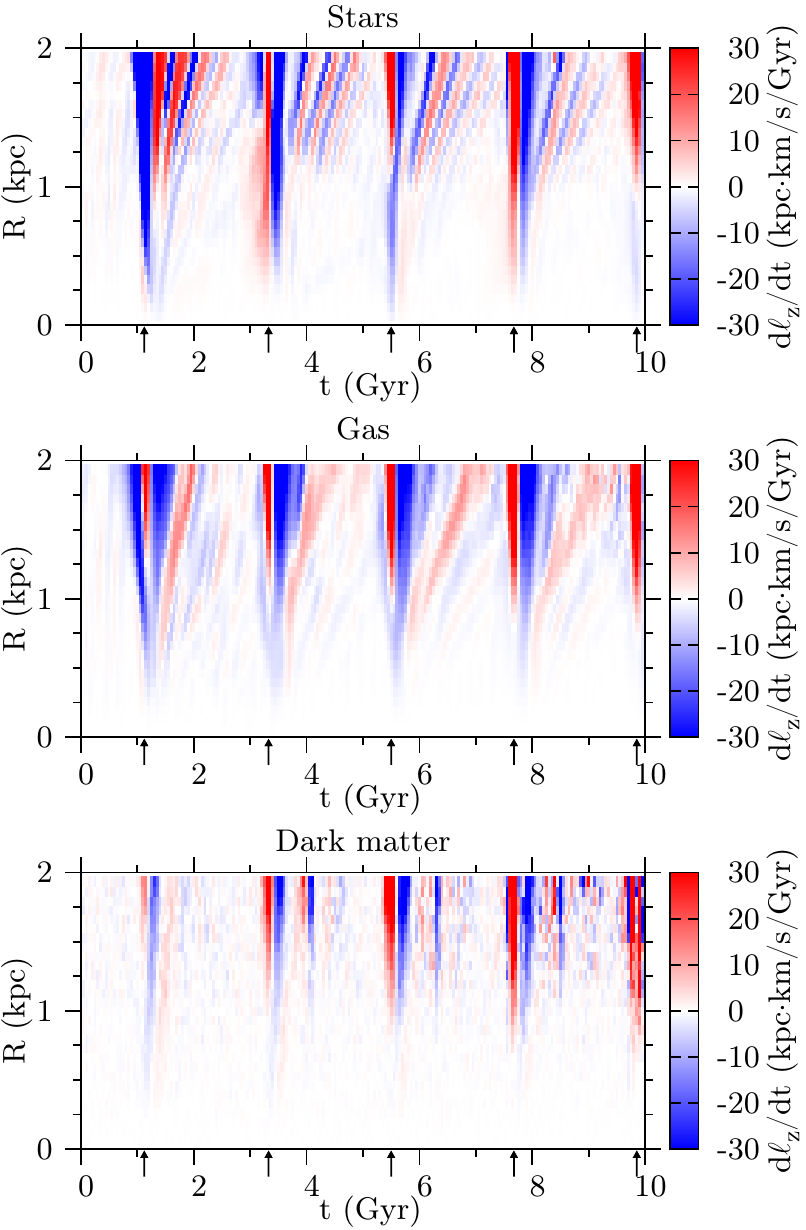}
\caption{Time derivative of the specific angular momentum component along the axis of rotation for the g30SF run. Top
panel: stars, middle panel: gas, bottom panel: dark matter. Small arrows below the panels indicate pericenter
passages.}
\label{fig_am_transfer}
\end{figure}

As we have already seen in Figure \ref{fig_a2_tot}, the bar in a~given galaxy can be strengthened or weakened at the
pericenter passage. To investigate the reason for such behavior, we measured the difference between the position angle
of the bar and the direction to the host galaxy, i.e.\ $\alpha=\varphi_\mathrm{bar}-\varphi_\mathrm{MW}$. The resulting
curves are shown in Figure \ref{fig_bar_position_angle}. The vertical scale covers the range from $-90^{\circ}$ to
$90^\circ$. Negative angles correspond to the bar lagging behind the line connecting the dwarf and the host and
positive ones mean that the bar has already overtaken the direction to the host. During their formation at the first
pericenter passage, the bars were aligned with the direction to the host galaxy and afterwards begun to rotate with
their own pattern speed. One can notice almost straight segments of the curves between the pericenters, which have
various slopes signifying different pattern speeds of the bars. This stems from the constant pattern speeds of the bars
between the pericenter passages while the orbital angular speed changed only a~little. In the vicinity of the
pericenters the orbital angular speed outpaced the bar rotation, resulting in the reversal of the slope of the curves.
The relative angle between the bar and the direction to the host galaxy is crucial in understanding the evolution of
the bar properties.

We illustrate our point further in Figure \ref{fig_peri_changes}. For each pericenter we calculated the value of the angle
$\alpha$. Then, we computed the differences of the pattern speed and of the bar strength between the preceding and following
apocenters. We did not include g70noSF, as its bar did not survive the second pericenter. For g70SF we took only
the second pericenter, as the bar was destroyed shortly after the third encounter with the host.
The anticorrelation of the changes in the bar strength and the pattern speed, as well as its dependence on the sign of $\alpha$, is clear.
Such an anticorrelation is expected as it was already found both in isolated and interacting systems \citep{athanassoula03, berentzen04}.
There is one green (g30noSF) point that does not fit the pattern. The bar in this model at the third pericenter was very weak and thus the angle might be slightly off.
One might expect that there should be extrema of changes at $\alpha=\pm45^\circ$ \citep{gerin90}.
However, recall that at each pericenter the dwarf galaxy subject to the tidal force was different.
If one performed a series of experiments changing only $\alpha$, a peak could appear.
However, here for each measurement not only the masses of the dwarf's components were different, but also their bar strengths and pattern speeds.
In particular, it is unclear whether $\Delta A_2$ would be the same if the bar had initially $|A_2|=0.2$ or $|A_2|=0.6$.
A similar dependence may occur for the pattern speed.
While the tidal force remains the same, the objects on which it acts do not.

We illustrate the action of the tidal force on the bar during the pericenters using the example of the Nbody run in
Figure \ref{fig_tidal_torque}. The tidal force acts approximately along the line connecting the center of the satellite
and the host. In the reference frame of the satellite, the tidal force stretches its body. However, when a barred
galaxy encounters the host, the tidal force effectively exerts a~torque on the bar and its effect depends on the bar
orientation with respect to the host at this time. During the second pericenter passage, depicted in the left panel of
the Figure, the tidal torque had the same orientation as the patter speed, which lead to the spinning up of the bar.
However, the stars at the tip of the bar were scattered and did not maintain coherent bar structure. Thus, the bar was
weakened and shortened. Next, during the third pericenter, the tidal torque had an opposite direction with respect to
the bar rotation. As a result, the bar was slowed down and was able to grow and gain more strength. Finally, at the
fourth pericenter, the situation was similar to the second pericenter so the bar was spun up and weakened.

The evolution of the dwarf in the g30SF run was qualitatively similar, with alternating weakening and strengthening of
the bar. In g30noSF, the bar was made stronger at the fourth pericenter, instead of being weakened. The evolution of
the gas-rich dwarfs was slightly more complicated, due to the destruction of their bars. In g70noSF the bar was already
in decline between the first and the second encounter with the host and was destroyed completely afterwards. In g70SF the
bar survived the second pericenter and was in fact enhanced by the tides, however it was declining afterwards.

Transfer of the angular momentum is usually considered as the main culprit responsible for the bar evolution. Thus, in
Figure \ref{fig_am_transfer} we plot the time derivative of the specific angular momentum component along the rotation
axis $\ell_z$. We include a~separate panel for each component of the g30SF dwarf. The evolution of this galaxy was
qualitatively similar to the Nbody run. The bar was weakened at the second and fourth pericenter passage and
strengthened at the third one.

The angular momentum of both rotating, baryonic components decreased at the first pericenter passage. Between the
encounters with the host there were some variations of the angular momentum in the bar region, but much weaker compared
to the bars formed through instability \citep[e.g.\ compare to][]{collier18}. However, the largest changes occurred
when the dwarf passed through the pericenter. During one such event, the stars both lost and gained angular momentum.
What was important was the net result. At the second and fourth pericenter the dwarf gained angular momentum, while at
the third one $\ell_z$ dropped.

The gaseous component lost angular momentum only during the pericenters. In galaxies that spontaneously create bars,
the dark matter halo absorbs angular momentum. Here, however, nothing similar happened. Actually, after the first
pericenter the halo had negative angular momentum due to preferential stripping of dark matter particles on prograde
orbits. When the halo has a prograde spin with respect to the disk it has a impact on the evolution of bars
\citep{saha_naab13, collier18}. \citet{saha_naab13} found that a halo spinning in a retrograde fashion slightly
suppresses the bar formation compared to a non-rotating halo. However, in our case the halo became slightly retrograde
at the same time as the bar formed, so it is difficult to judge its influence.

\subsection{Star formation}

We now move to the analysis of the star formation processes in dwarf galaxies.
Of course, in our simulations we did not have the full grasp of all processes that influence the star formation.
Firstly, we did not include the gaseous halos of both galaxies.
The hot gas halo of the host would have induced ram pressure stripping of the dwarf's gas. The gas in the halo of the dwarf could cool and fall down onto the dwarf's disk.
Secondly, the star formation model we used in our simulations was developed for larger galaxies at lower resolution, thus it may not encompass all relevant physics.

\begin{figure}
\centering
\includegraphics{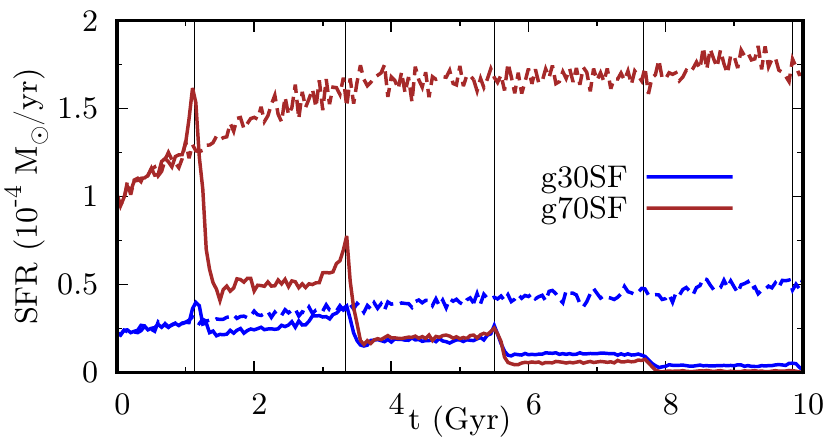}
\caption{Star formation rates of the dwarfs orbiting the host galaxy (solid lines) compared to the same galaxies
evolved in isolation (dashed lines).
Vertical lines indicate pericenter passages of the satellites.}
\label{fig_sfr}
\end{figure}

In Figure \ref{fig_sfr} we plot the evolution of the star formation rate (SFR) in the inner $2$ kpc for both g30SF and
g70SF. For comparison, we also add data for the same dwarfs evolved in isolation. In the dwarfs orbiting the host, SFR
was enhanced during the first pericenter passage by about $30\%$. This was a result of gas compression by the tidal
force in the central part of the dwarfs. Later, it dropped significantly due to stripping of the gas and thickening of
the gas distribution, which reduced its volume density in the central parts of the galaxies. During the following
orbital period, SFR was constant or slightly increasing. At the next pericenters it was enhanced again and dropped
afterwards. Interestingly, at the final stages of the evolution, SFR of the initially more gas rich dwarf was lower.
While the total gas content of g70SF was higher than in g30SF at all times, its distribution was different, in
particular it was thicker.

\begin{figure}
	\centering
	\includegraphics{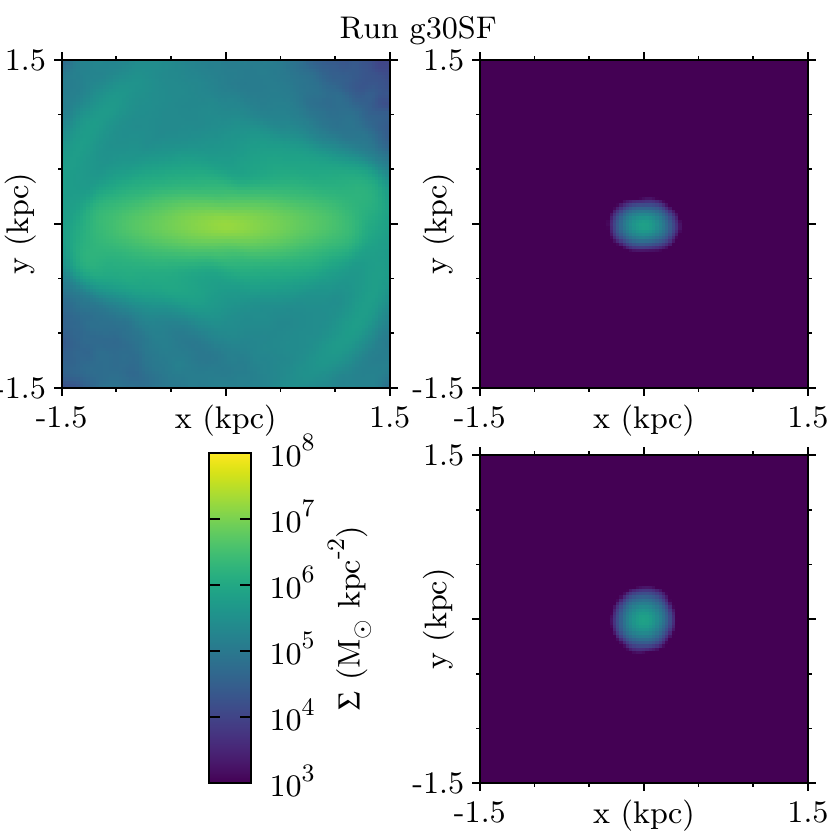}
	\includegraphics{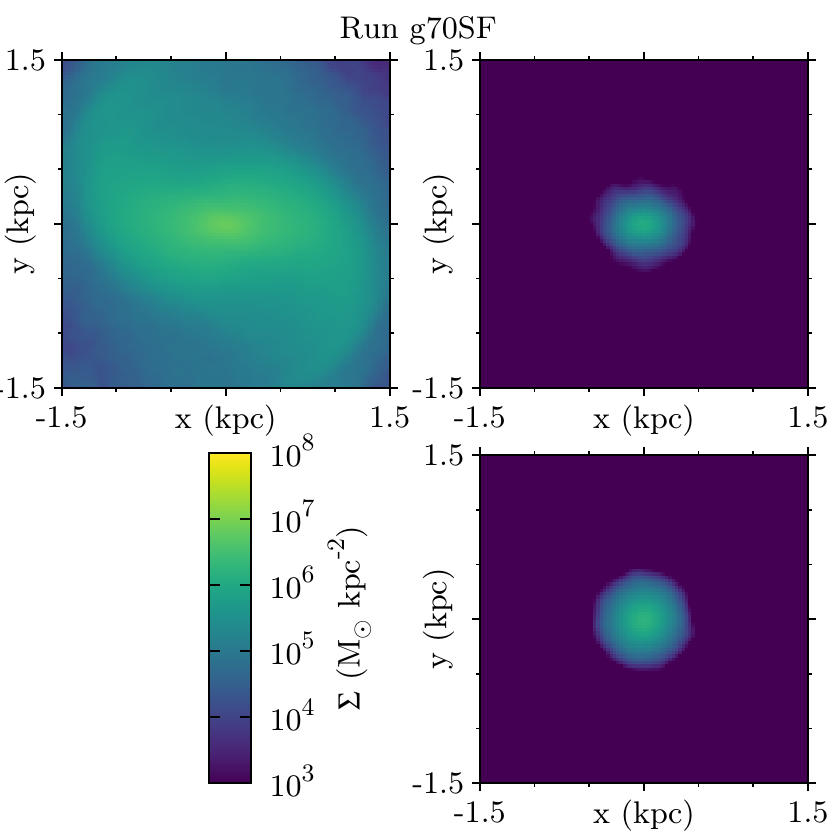}
	\caption{Each triplet of plots illustrates the surface densities of stars in one simulation. Top left panels of
	each triplet show the
	distribution of old stars at the second apocenter ($t=2.2$ Gyr). Top right panels plot the distribution of new
	stars formed between the first pericenter ($t=1.125$ Gyr) and the second apocenter. Bottom right panels show the
	distribution of new stars formed during the corresponding time in the isolated dwarf.}
\label{fig_new_stars}
\end{figure}

In Figure \ref{fig_new_stars} we compare the distribution of old stars, present in the simulation at the beginning, and
new stars formed after the first pericenter passage. Both the isolated and the satellite dwarf formed stars in their
centermost part. However, in the satellite galaxy there were obviously less stars formed and their distribution was
elongated in the same direction as the bar made of the old stars. We inspected the time evolution of the newly formed
stars and it turned out that the elongation rotated with the same pattern speed as the bar.

We note that in the observed dwarf irregulars the star formation is not concentrated in their central parts,
but it is more spread and takes place in knots \citep[e.g.\ ][]{bianchi12}. We believe that the mismatch between our
results and observations is due to the simplified model of star formation in our simulations.

\section{Discussion}
\label{sec_discussion}
\subsection{Gas behavior in the central region}

It is interesting to compare our results regarding tidally induced bars with the more established knowledge regarding
bars created in unstable disks. In this work we focused on the impact of the gaseous component. In previous works it was
found that gas inhibits the instability in disk galaxies and the resulting bars are shorter and weaker
\citep{berentzen07, athanassoula13gas}. The bars induced in our simulations were indeed weaker for a higher gas
fraction, but they had approximately the same lengths.

The overall behavior of the gaseous component is the most pronounced difference. Simulations of gas flow in barred
galaxies, both these performed in rigid potentials \citep{binney91, athanassoula92b, kim12, sormani15} and the fully
self-consistent ones \citep[e.g.\ ][]{athanassoula13gas, pettitt_wadsley18}, all lead to very similar conclusions. The
gas is driven toward the center of a galaxy, which leads to the formation of a central mass concentration (CMC)
surrounded by a region practically devoid of gas, only penetrated by off-centered shocks, located at the leading sides
of the bar. The largest inflow of gas takes place right at the moment of the bar formation and then continues, but at a
slower rate \citep{villa_vargas10, athanassoula13gas}. In the simulations presented here the behavior of gas was
entirely different. Its distribution remained approximately axisymmetric and did not exhibit any net inflow. Some
rearrangement of the gaseous material took place, but not on such a scale as in previous works.

We can try to point out some differences of our model with respect to previous work. Obviously, our bars were tidally
induced, whereas previously they were either adiabatically grown in a given potential or formed through inherent
instability. \citet{pettitt_wadsley18} also studied the formation of tidally induced bars in their simulations and
their galaxies exhibited the usual pattern of the gas flow. However, their models were more massive and bar-unstable
even in isolation, so the interaction only changed how fast the bars formed or their final strength. We explicitly
checked that our models were stable against bar creation in isolation for times comparable to the whole length of the
discussed runs.

Another potentially important difference is due to the amount of gas, as our gas fractions are very high. For example,
\citet{pettitt_wadsley18} used only a $10\%$ gas fraction and while in \citet{athanassoula13gas} in some runs the
galaxies were clearly dominated by gas initially, it was quickly consumed by star formation and at the epoch of the bar
formation the baryonic component was already dominated by stars. Such a low fraction of gas corresponds to real values
observed in Milky Way-sized galaxies. However, dwarf galaxies are much more gas-rich.

Other possible explanation could be that our bars were abruptly created at the first pericenter passage.
In the cosmological zoom-in simulation of \citet{spinoso17} the bar did not drive the gas toward the center immediately and needed a few Gyr and a few rotations to do so.
On the other hand, in the simulation of \cite{kim12}, the adiabatically grown potential of the bar alters the gas flow in about half a rotation.
Our bars, after their formation at the first pericenter passage, experienced only about $1.5$ rotations before
the next encounter with the host, so one could think that they did not have enough time to sweep the gas. To verify if
this was the case we performed the following experiment. We took the g30noSF dwarf at the time of its second apocenter
($2.2$ Gyr), placed it in isolation and evolved for $7.8$ Gyr. During this time the gas was not driven to the center of
the dwarf. The bar itself did not evolve significantly either.
Our bars are clearly rotating slowly, in terms of
$\mathcal{R}$, but some of the bars of \citet{spinoso17} and \citet{pettitt_wadsley18} were also slow.

In the view of \citet{binney91} and \citet{athanassoula92b} the creation of shocks is strongly linked to the properties
of the underlying periodic orbits. Both these studies emphasized the importance of loops or cusps at the apocenters of
the x$_1$ orbits. As the gas follows periodic orbits, it should be shocked at cusps and loops, because gas is
collisional and its streamlines cannot cross. Such cusps or loops, however, are located at the bar major axis, while
the shock loci are in most cases displaced to the leading side of the bar. \citet{athanassoula92b} showed that this can
be achieved if the x2 and x3 families not only exist, but also extend sufficiently far from the center. We did not
fully analyze the orbital structure of our models, however in \citet{gajda16}, in a similar setting, we found some
near-periodic x$_1$ orbits with loops at their ends, so the absence of such orbits cannot be responsible.

Close inspection of simulations by \citet{athanassoula92b} and \citet{athanassoula13gas} reveals that the size of their
gas-deficient region depends on the strength of the bar. While values of $|A_2|$ of our bars are not low (e.g.\
$|A_{2\mathrm{, max}}|\sim0.6$), our simulations lacked other signatures of bar strength, such as rings and ansae.
Furthermore, the dwarfs were dominated by their dark matter haloes, further diminishing the influence of the bar.

The temperature and the equation of state of the gas also influence its behavior. In this work we
reported on simulations that followed either ideal gas undergoing adiabatic evolution or a stiff equation of state from
\citet{springel_hernquist03}. We performed a few additional experiments with model g30noSF, but at a particle resolution
four times lower than in the runs presented in this work. Firstly, we changed the initial temperature of the gas under
ideal gas law. Instead of $2000$ K we tried both $20\ 000$ K and $200$ K. At the higher temperature there was even less
response in the gaseous component. At the lower initial temperature there was slightly more response to the
perturbation, but the gas was heated during the interaction to the similar temperature as in the runs with a higher
initial temperature.

We also checked the impact of using the isothermal equation of state.
As this implies an adiabatic index of $\gamma=1$, there was no generation of heat due to artificial viscosity, which is introduced in SPH to accurately capture shocks
and discontinuities \citep{springel05}. 
In our test we used a sound speed equivalent of $2000$ K.
After the bar was formed at the first pericenter passage, the gas distribution followed the bar shape and subsequently it
was driven to the center of the dwarf, leading to formation of a CMC.
It seems that if the isothermal equation of state is used, the pressure response to changes in the density is smaller, which enables such a behavior.
One should not conclude that the equation of state is the only factor responsible for the gas unresponsiveness.
\citet{athanassoula13gas} used the same equation of state as we did in the star forming runs.
\citet{villa_vargas10} and \citet{pettitt_wadsley18} used isothermal gas of $T=10^4$ K, which is higher than the values reached in our runs, hence their gas was at relatively higher pressures than in our case. 
In these works the gas was driven towards the centers of galaxies.
Perhaps the gas response depends on the balance between non-axisymmetric driving and pressure increase due to compression.
The study of the impact of the equation of state on the tidally induced bar formation is beyond the scope of this work and should be addressed elsewhere.

Interestingly, the only dwarf galaxy in the Local Group which unquestionably possesses a bar, namely the Large
Magellanic Cloud, also has an undisturbed distribution of HI gas \citep{staveley-smith03}. Note however, that this bar
probably was not induced by the Milky Way because the galaxy is likely on its first approach to the Milky Way
\citep{besla07, besla12} and the main culprit for its current state is the interaction with the Small Magellanic
Cloud \citep{pardy16}. It turns out that in many Magellanic-type galaxies the bar seems to disturb significantly
neither the distribution nor the kinematics of the gaseous component \citep{wilcots09}.

\subsection{Tidal torques}

The effect of tidal torques on already existing bars was already studied by \citet{gerin90}. They investigated the
response of a bar created via instability to a perturbation. They found that the bar strength was briefly enhanced or
reduced, depending on the relative orientation of the bar and the perturber at the pericenter. Later, their bars
returned to their original amplitude. In our case the bars were permanently altered, presumably because their model was
unstable while ours was stable. They conjectured that the strongest influence takes place when the relative angle is
$\alpha=\pm 45^{\circ}$. The study was later expanded by \citet{sundin_sundelius91} and \citet{sundin93} who found that
the aforementioned effect occurs only for strong interactions. In a case of a weak interaction, the stellar component
always loses angular momentum, however, the impact on the bar strength was not analyzed.

In this work we confirmed these previous results, but with much more reliable simulations. Due to nearly 30 years
of improvements of computer technology, we were able to use two orders of magnitude more particles, we studied gas-rich
bars and witnessed the mechanism working at multiple pericenter passages. \citet{lokas14} also demonstrated that tidal
torques influence the bar pattern speed. However, they studied this behavior in detail only during one pericenter
passage, at which the bar was first slowed down and then speeded up. Here, we witnessed that a similar thing happens at
each pericenter passage, as revealed by the angular momentum changes, but the final outcome of the encounter depends on
the integrated effect.

\subsection{Bar survival}

Bars in the runs with small amount of gas ($0\%$ and $30\%$) survived until the end of the simulations. However, in the
gas-rich runs the bars were destroyed relatively quickly and were not regenerated later. It was reported in the past
that bars in simulations including the gaseous component can be severely weakened or even destroyed
\citep{bournaud_combes02, berentzen04, bournaud05}. This was initially ascribed to the build-up of the CMC and the gas
torque action on the stellar component. However, later works put forward a different interpretation \citep{berentzen07,
athanassoula13gas} and argued that it was due to buckling coupled to the use of rigid haloes.

In our runs we did not witness extensive buckling nor substantial growth of a CMC. However, while the total mass inside
$0.5$ kpc remained roughly constant in all the runs, some rearrangement of the material took place. The baryonic
mass inside central $0.1$ kpc in all runs grew on average by a factor of $1.6$, which in the high gas fraction run could
have been enough to induce scattering of the stellar particles. Additionally, in the high-gas fraction runs the impact
of the stellar component on the potential was smaller than the one of the gas, so the stellar particles could diffuse
more easily to orbits which did not support the bar.

\subsection{Thickening and buckling}

Buckling is a common phenomenon in the evolution of barred galaxies \citep{combes_sanders81, raha91}, which leads to
formation of boxy/peanut (b/p) bulges \citep[see][for a review]{athanassoula16}. We checked if any such events took
place in our dwarf galaxies. After careful examination, we have found that in the Nbody run the bar buckled just before
the second pericenter passage. However, no b/p bulge has formed, probably because of two reasons. Firstly, the dwarf
was shocked right afterwards by the tides, so its structure was disturbed. Secondly, the outer parts of the disk also
thickened, so the b/p bulge could not be discerned. We did not find the other, gaseous runs to experience buckling.
However, they swelled just after their bars formed at the first pericenter. This is in line with the results of
\citet{berentzen07} who found that bars in gas-rich galaxies do not buckle, but thicken soon after the formation of the
bar. They ascribed this phenomenon to the vertical scattering of orbits by the CMC. However, no significant CMCs formed
in our dwarfs. Further work is thus necessary to achieve a better understanding of the thickening and buckling of
the inner bar regions.

\subsection{Comparison with similar studies}

Recently, \citet{kazantzidis17} followed the evolution of dwarf galaxies in a setup similar to ours, however they
included a hot gas halo of the host galaxy, which leads to the ram pressure stripping of the satellite's gas. In
addition, their dwarf was an order of magnitude more massive and they employed vigorous supernova feedback. While they
reported that no bar was created, this conclusion may have been influenced by the way the bar mode was measured.
Namely, it was computed inside only one radial disk scale-length and the authors assumed that a bar was created only if
$|A_2|>0.2$. If we had measured the bar strength inside our initial disk scale-length (i.e.\ $0.41$ kpc), we would also
have not found bars. Comparing our results to \citet{kazantzidis17} one has to remember that severe stripping of the
gas may significantly alter the dynamics of the disk.

\section{Conclusions}
\label{sec_conclusions}

In this work we studied the impact of the interstellar medium on tidally induced bars in dwarf galaxies orbiting a
Milky Way-like host. To investigate this issue we used $N$-body/SPH simulations with various amounts of gas. We
performed runs with $0\%$, $30\%$ and $70\%$ gas fractions. The simulations with gas were performed twice. In one
version the star formation processes were not included, while in the other we turned them on. All other parameters of
the simulations, such as the masses of the galaxies or the orbital parameters, were kept the same.

In all our simulations, bars formed at the first pericenter passage. Despite different gas fractions, the bars had
initially the same length and pattern speed. In the gas-poor runs ($0\%$ and $30\%$) further evolution was governed by
the action of tidal torques at subsequent pericenters. If the bar was lagging behind the host it was spun up and
shortened, while if the bar was leading it was slowed down and elongated. The evolution of the galaxies in the gas-rich
runs ($70\%$) was different. The strength of their bars decreased between pericenters leading to their complete
destruction.

The bars formed in these simulations had a number of features that distinguish them from most of the simulated bars
formed in isolation. Our bars did not grow steadily like bars in the secular phase. They also did not slow down through
the exchange of angular momentum with outer parts of the disk and the dark matter halo. Their ratios of the corotation
radius to the bar length, $\mathcal{R}_\mathrm{cr}=2$--$3$, place them among slow bars. These results are similar to
what we found in \citet{gajda17} and what was reported in the past, by e.g.\ \citet{miwa_noguchi98}. However, slow bars
can be created spontaneously if the disk is dark matter dominated and there is little or no gas
\citep{debattista_sellwood00, saha_naab13, athanassoula14, lokas16, chequers16, pettitt_wadsley18}. Our dwarfs were also
dominated by dark matter.

The main obvious drawback of our study is the lack of the hot gas halo of the Milky Way. We did not include it, as we
wanted to focus on the internal dynamics of the dwarfs and the inclusion of the gas halo would require us to severely
reduce resolution. The action of the gaseous halo leads to more stripping of the gas from the dwarf \citep{mayer06,
kazantzidis17, simpson18}, which could in turn alter the dynamics of the bar. Although the gas is probably efficiently
ram-pressure stripped by the hot halo of the host, some of it may remain in the dwarf long enough to significantly
affect the evolution. If the bar destruction in our gas rich runs was caused by the gaseous component, then its
stripping could have allowed the bars to survive for much longer.

Another issue that hindered our understanding of the bar dynamics were the repeated tidal interactions with the host.
While the two processes driven by the tidal interaction, namely the bar creation and its further encounters with the
host, are very interesting on their own, their study was difficult. Very little time passed between pericenter
passages, thus we were not able to detect any long-term (secular) trends, as they were interrupted. We have chosen such
orbits in order to investigate the evolution of satellite dwarf galaxies, but a study focused on the physics of bars
would also benefit from a different model, in which there is only a single interaction, as it is usually done
\citep[e.g.\ ][]{gerin90, miwa_noguchi98, pettitt_wadsley18, lokas18}.

\acknowledgments
We would like to thank V. Springel for providing us a copy of and permission to use \textsc{Gadget3}.
This work was partially supported by the Polish National Science Centre under grant 2013/10/A/ST9/00023.
GG is supported by the Foundation for Polish Science (FNP) within the START program.
EA is partially supported by the CNES (Centre National d'\'Etudes Spatiales, France).
We would like to thank the anonymous referee for their constructive feedback.


\begin{thebibliography}{}
\expandafter\ifx\csname natexlab\endcsname\relax\def\natexlab#1{#1}\fi
\providecommand{\url}[1]{\href{#1}{#1}}

\bibitem[{{Aguerri} \&
  {Gonz{\'a}lez-Garc{\'{\i}}a}(2009)}]{aguerri_gonzalez-garcia09}
{Aguerri}, J.~A.~L., \& {Gonz{\'a}lez-Garc{\'{\i}}a}, A.~C. 2009, \aap, 494,
  891

\bibitem[{{Algorry} {et~al.}(2017){Algorry}, {Navarro}, {Abadi}, {Sales},
  {Bower}, {Crain}, {Dalla Vecchia}, {Frenk}, {Schaller}, {Schaye}, \&
  {Theuns}}]{algorry17}
{Algorry}, D.~G., {Navarro}, J.~F., {Abadi}, M.~G., {et~al.} 2017, \mnras, 469,
  1054

\bibitem[{{Athanassoula}(1980)}]{athanassoula80}
{Athanassoula}, E. 1980, \aap, 88, 184

\bibitem[{{Athanassoula}(1992{\natexlab{a}})}]{athanassoula92a}
---. 1992{\natexlab{a}}, \mnras, 259, 328

\bibitem[{{Athanassoula}(1992{\natexlab{b}})}]{athanassoula92b}
---. 1992{\natexlab{b}}, \mnras, 259, 345

\bibitem[{{Athanassoula}(2002)}]{athanassoula02}
---. 2002, \apjl, 569, L83

\bibitem[{{Athanassoula}(2003)}]{athanassoula03}
---. 2003, \mnras, 341, 1179

\bibitem[{{Athanassoula}(2013)}]{athanassoula13review}
---. 2013, {Bars and secular evolution in disk galaxies: Theoretical input},
  ed. J.~{Falc{\'o}n-Barroso} \& J.~H. {Knapen}, 305

\bibitem[{{Athanassoula}(2014)}]{athanassoula14}
---. 2014, \mnras, 438, L81

\bibitem[{{Athanassoula}(2016)}]{athanassoula16}
{Athanassoula}, E. 2016, in Astrophysics and Space Science Library, Vol. 418,
  Galactic Bulges, ed. E.~{Laurikainen}, R.~{Peletier}, \& D.~{Gadotti}, 391

\bibitem[{{Athanassoula} {et~al.}(2013){Athanassoula}, {Machado}, \&
  {Rodionov}}]{athanassoula13gas}
{Athanassoula}, E., {Machado}, R.~E.~G., \& {Rodionov}, S.~A. 2013, \mnras,
  429, 1949

\bibitem[{{Athanassoula} \& {Misiriotis}(2002)}]{athanassoula_misiriotis02}
{Athanassoula}, E., \& {Misiriotis}, A. 2002, \mnras, 330, 35

\bibitem[{{Ben{\'{\i}}tez-Llambay} {et~al.}(2018){Ben{\'{\i}}tez-Llambay},
  {Navarro}, {Frenk}, \& {Ludlow}}]{benitez-llambay18}
{Ben{\'{\i}}tez-Llambay}, A., {Navarro}, J.~F., {Frenk}, C.~S., \& {Ludlow},
  A.~D. 2018, \mnras, 473, 1019

\bibitem[{{Berentzen} {et~al.}(2004){Berentzen}, {Athanassoula}, {Heller}, \&
  {Fricke}}]{berentzen04}
{Berentzen}, I., {Athanassoula}, E., {Heller}, C.~H., \& {Fricke}, K.~J. 2004,
  \mnras, 347, 220

\bibitem[{{Berentzen} {et~al.}(1998){Berentzen}, {Heller}, {Shlosman}, \&
  {Fricke}}]{berentzen98}
{Berentzen}, I., {Heller}, C.~H., {Shlosman}, I., \& {Fricke}, K.~J. 1998,
  \mnras, 300, 49

\bibitem[{{Berentzen} {et~al.}(2007){Berentzen}, {Shlosman},
  {Martinez-Valpuesta}, \& {Heller}}]{berentzen07}
{Berentzen}, I., {Shlosman}, I., {Martinez-Valpuesta}, I., \& {Heller}, C.~H.
  2007, \apj, 666, 189

\bibitem[{{Besla} {et~al.}(2007){Besla}, {Kallivayalil}, {Hernquist},
  {Robertson}, {Cox}, {van der Marel}, \& {Alcock}}]{besla07}
{Besla}, G., {Kallivayalil}, N., {Hernquist}, L., {et~al.} 2007, \apj, 668, 949

\bibitem[{{Besla} {et~al.}(2012){Besla}, {Kallivayalil}, {Hernquist}, {van der
  Marel}, {Cox}, \& {Kere{\v s}}}]{besla12}
---. 2012, \mnras, 421, 2109

\bibitem[{{Bianchi} {et~al.}(2012){Bianchi}, {Efremova}, {Hodge}, {Massey}, \&
  {Olsen}}]{bianchi12}
{Bianchi}, L., {Efremova}, B., {Hodge}, P., {Massey}, P., \& {Olsen}, K.~A.~G.
  2012, \aj, 143, 74

\bibitem[{{Binney} {et~al.}(1991){Binney}, {Gerhard}, {Stark}, {Bally}, \&
  {Uchida}}]{binney91}
{Binney}, J., {Gerhard}, O.~E., {Stark}, A.~A., {Bally}, J., \& {Uchida}, K.~I.
  1991, \mnras, 252, 210

\bibitem[{{Bournaud} \& {Combes}(2002)}]{bournaud_combes02}
{Bournaud}, F., \& {Combes}, F. 2002, \aap, 392, 83

\bibitem[{{Bournaud} {et~al.}(2005){Bournaud}, {Combes}, \&
  {Semelin}}]{bournaud05}
{Bournaud}, F., {Combes}, F., \& {Semelin}, B. 2005, \mnras, 364, L18

\bibitem[{{Chandrasekhar}(1943)}]{chandrasekhar43}
{Chandrasekhar}, S. 1943, \apj, 97, 255

\bibitem[{{Chequers} {et~al.}(2016){Chequers}, {Spekkens}, {Widrow}, \&
  {Gilhuly}}]{chequers16}
{Chequers}, M.~H., {Spekkens}, K., {Widrow}, L.~M., \& {Gilhuly}, C. 2016,
  \mnras, 463, 1751

\bibitem[{{Cheung} {et~al.}(2013){Cheung}, {Athanassoula}, {Masters}, {Nichol},
  {Bosma}, {Bell}, {Faber}, {Koo}, {Lintott}, {Melvin}, {Schawinski}, {Skibba},
  \& {Willett}}]{cheung13}
{Cheung}, E., {Athanassoula}, E., {Masters}, K.~L., {et~al.} 2013, \apj, 779,
  162

\bibitem[{{Coleman} {et~al.}(2007){Coleman}, {de Jong}, {Martin}, {Rix},
  {Sand}, {Bell}, {Pogge}, {Thompson}, {Hippelein}, {Giallongo}, {Ragazzoni},
  {DiPaola}, {Farinato}, {Smareglia}, {Testa}, {Bechtold}, {Hill}, {Garnavich},
  \& {Green}}]{coleman07}
{Coleman}, M.~G., {de Jong}, J.~T.~A., {Martin}, N.~F., {et~al.} 2007, \apjl,
  668, L43

\bibitem[{{Collier} {et~al.}(2018){Collier}, {Shlosman}, \&
  {Heller}}]{collier18}
{Collier}, A., {Shlosman}, I., \& {Heller}, C. 2018, \mnras, 476, 1331

\bibitem[{{Combes} {et~al.}(1990){Combes}, {Debbasch}, {Friedli}, \&
  {Pfenniger}}]{combes90}
{Combes}, F., {Debbasch}, F., {Friedli}, D., \& {Pfenniger}, D. 1990, \aap,
  233, 82

\bibitem[{{Combes} \& {Sanders}(1981)}]{combes_sanders81}
{Combes}, F., \& {Sanders}, R.~H. 1981, \aap, 96, 164

\bibitem[{{Contopoulos}(1980)}]{contopoulos80}
{Contopoulos}, G. 1980, \aap, 81, 198

\bibitem[{{Corsini}(2011)}]{corsini11}
{Corsini}, E.~M. 2011, Memorie della Societa Astronomica Italiana Supplementi,
  18, 23

\bibitem[{{Debattista} {et~al.}(2006){Debattista}, {Mayer}, {Carollo}, {Moore},
  {Wadsley}, \& {Quinn}}]{debattista06}
{Debattista}, V.~P., {Mayer}, L., {Carollo}, C.~M., {et~al.} 2006, \apj, 645,
  209

\bibitem[{{Debattista} \& {Sellwood}(2000)}]{debattista_sellwood00}
{Debattista}, V.~P., \& {Sellwood}, J.~A. 2000, \apj, 543, 704

\bibitem[{{Dubinski} {et~al.}(2009){Dubinski}, {Berentzen}, \&
  {Shlosman}}]{dubinski09}
{Dubinski}, J., {Berentzen}, I., \& {Shlosman}, I. 2009, \apj, 697, 293

\bibitem[{{Erwin}(2018)}]{erwin18}
{Erwin}, P. 2018, \mnras, 474, 5372

\bibitem[{{Erwin} \& {Debattista}(2017)}]{erwin_debattista17}
{Erwin}, P., \& {Debattista}, V.~P. 2017, \mnras, 468, 2058

\bibitem[{{Fabrizio} {et~al.}(2016){Fabrizio}, {Bono}, {Nonino}, {{\L}okas},
  {Ferraro}, {Iannicola}, {Buonanno}, {Cassisi}, {Coppola}, {Dall'Ora},
  {Gilmozzi}, {Marconi}, {Monelli}, {Romaniello}, {Stetson}, {Th{\'e}venin}, \&
  {Walker}}]{fabrizio16}
{Fabrizio}, M., {Bono}, G., {Nonino}, M., {et~al.} 2016, \apj, 830, 126

\bibitem[{{Font} {et~al.}(2017){Font}, {Beckman}, {Mart{\'{\i}}nez-Valpuesta},
  {Borlaff}, {James}, {D{\'{\i}}az-Garc{\'{\i}}a}, {Garc{\'{\i}}a-Lorenzo},
  {Camps-Fari{\~n}a}, {Guti{\'e}rrez}, \& {Amram}}]{font17}
{Font}, J., {Beckman}, J.~E., {Mart{\'{\i}}nez-Valpuesta}, I., {et~al.} 2017,
  \apj, 835, 279

\bibitem[{{Frings} {et~al.}(2017){Frings}, {Macci{\`o}}, {Buck}, {Penzo},
  {Dutton}, {Blank}, \& {Obreja}}]{frings17}
{Frings}, J., {Macci{\`o}}, A., {Buck}, T., {et~al.} 2017, \mnras, 472, 3378

\bibitem[{{Gabbasov} {et~al.}(2006){Gabbasov}, {Rodr{\'{\i}}guez-Meza},
  {Klapp}, \& {Cervantes-Cota}}]{gabbasov06}
{Gabbasov}, R.~F., {Rodr{\'{\i}}guez-Meza}, M.~A., {Klapp}, J., \&
  {Cervantes-Cota}, J.~L. 2006, \aap, 449, 1043

\bibitem[{{Gajda} {et~al.}(2016){Gajda}, {{\L}okas}, \&
  {Athanassoula}}]{gajda16}
{Gajda}, G., {{\L}okas}, E.~L., \& {Athanassoula}, E. 2016, \apj, 830, 108

\bibitem[{{Gajda} {et~al.}(2017){Gajda}, {{\L}okas}, \&
  {Athanassoula}}]{gajda17}
---. 2017, \apj, 842, 56

\bibitem[{{Gajda} {et~al.}(2015){Gajda}, {{\L}okas}, \& {Wojtak}}]{gajda15}
{Gajda}, G., {{\L}okas}, E.~L., \& {Wojtak}, R. 2015, \mnras, 447, 97

\bibitem[{{Gerin} {et~al.}(1990){Gerin}, {Combes}, \& {Athanassoula}}]{gerin90}
{Gerin}, M., {Combes}, F., \& {Athanassoula}, E. 1990, \aap, 230, 37

\bibitem[{{Gingold} \& {Monaghan}(1977)}]{gingold_monaghan77}
{Gingold}, R.~A., \& {Monaghan}, J.~J. 1977, \mnras, 181, 375

\bibitem[{{Hohl}(1971)}]{hohl71}
{Hohl}, F. 1971, \apj, 168, 343

\bibitem[{{Janz} {et~al.}(2012){Janz}, {Laurikainen}, {Lisker}, {Salo},
  {Peletier}, {Niemi}, {den Brok}, {Toloba}, {Falc{\'o}n-Barroso}, {Boselli},
  \& {Hensler}}]{janz12}
{Janz}, J., {Laurikainen}, E., {Lisker}, T., {et~al.} 2012, \apjl, 745, L24

\bibitem[{{Katz} {et~al.}(1996){Katz}, {Weinberg}, \& {Hernquist}}]{katz96}
{Katz}, N., {Weinberg}, D.~H., \& {Hernquist}, L. 1996, \apjs, 105, 19

\bibitem[{{Kazantzidis} {et~al.}(2011){Kazantzidis}, {{\L}okas}, {Callegari},
  {Mayer}, \& {Moustakas}}]{kazantzidis11}
{Kazantzidis}, S., {{\L}okas}, E.~L., {Callegari}, S., {Mayer}, L., \&
  {Moustakas}, L.~A. 2011, \apj, 726, 98

\bibitem[{{Kazantzidis} {et~al.}(2017){Kazantzidis}, {Mayer}, {Callegari},
  {Dotti}, \& {Moustakas}}]{kazantzidis17}
{Kazantzidis}, S., {Mayer}, L., {Callegari}, S., {Dotti}, M., \& {Moustakas},
  L.~A. 2017, \apjl, 836, L13

\bibitem[{{Kim} {et~al.}(2012){Kim}, {Seo}, {Stone}, {Yoon}, \&
  {Teuben}}]{kim12}
{Kim}, W.-T., {Seo}, W.-Y., {Stone}, J.~M., {Yoon}, D., \& {Teuben}, P.~J.
  2012, \apj, 747, 60

\bibitem[{{Klimentowski} {et~al.}(2009){Klimentowski}, {{\L}okas},
  {Kazantzidis}, {Mayer}, \& {Mamon}}]{klimentowski09}
{Klimentowski}, J., {{\L}okas}, E.~L., {Kazantzidis}, S., {Mayer}, L., \&
  {Mamon}, G.~A. 2009, \mnras, 397, 2015

\bibitem[{{Kraljic} {et~al.}(2012){Kraljic}, {Bournaud}, \&
  {Martig}}]{kraljic12}
{Kraljic}, K., {Bournaud}, F., \& {Martig}, M. 2012, \apj, 757, 60

\bibitem[{{Lang} {et~al.}(2014){Lang}, {Holley-Bockelmann}, \&
  {Sinha}}]{lang14}
{Lang}, M., {Holley-Bockelmann}, K., \& {Sinha}, M. 2014, \apjl, 790, L33

\bibitem[{{Lee} {et~al.}(2012){Lee}, {Park}, {Lee}, \& {Choi}}]{lee12}
{Lee}, G.-H., {Park}, C., {Lee}, M.~G., \& {Choi}, Y.-Y. 2012, \apj, 745, 125

\bibitem[{{{\L}okas}(2018)}]{lokas18}
{{\L}okas}, E.~L. 2018, \apj, 857, 6

\bibitem[{{{\L}okas} {et~al.}(2014){{\L}okas}, {Athanassoula}, {Debattista},
  {Valluri}, {Pino}, {Semczuk}, {Gajda}, \& {Kowalczyk}}]{lokas14}
{{\L}okas}, E.~L., {Athanassoula}, E., {Debattista}, V.~P., {et~al.} 2014,
  \mnras, 445, 1339

\bibitem[{{{\L}okas} {et~al.}(2016){{\L}okas}, {Ebrov{\'a}}, {del Pino},
  {Sybilska}, {Athanassoula}, {Semczuk}, {Gajda}, \& {Fouquet}}]{lokas16}
{{\L}okas}, E.~L., {Ebrov{\'a}}, I., {del Pino}, A., {et~al.} 2016, \apj, 826,
  227

\bibitem[{{{\L}okas} {et~al.}(2010){{\L}okas}, {Kazantzidis}, {Majewski},
  {Law}, {Mayer}, \& {Frinchaboy}}]{lokas10}
{{\L}okas}, E.~L., {Kazantzidis}, S., {Majewski}, S.~R., {et~al.} 2010, \apj,
  725, 1516

\bibitem[{{{\L}okas} {et~al.}(2011){{\L}okas}, {Kazantzidis}, \&
  {Mayer}}]{lokas11}
{{\L}okas}, E.~L., {Kazantzidis}, S., \& {Mayer}, L. 2011, \apj, 739, 46

\bibitem[{{{\L}okas} {et~al.}(2012){{\L}okas}, {Majewski}, {Kazantzidis},
  {Mayer}, {Carlin}, {Nidever}, \& {Moustakas}}]{lokas12}
{{\L}okas}, E.~L., {Majewski}, S.~R., {Kazantzidis}, S., {et~al.} 2012, \apj,
  751, 61

\bibitem[{{{\L}okas} {et~al.}(2015){{\L}okas}, {Semczuk}, {Gajda}, \&
  {D'Onghia}}]{lokas15}
{{\L}okas}, E.~L., {Semczuk}, M., {Gajda}, G., \& {D'Onghia}, E. 2015, \apj,
  810, 100

\bibitem[{{Lucy}(1977)}]{lucy77}
{Lucy}, L.~B. 1977, \aj, 82, 1013

\bibitem[{{Masters} {et~al.}(2011){Masters}, {Nichol}, {Hoyle}, {Lintott},
  {Bamford}, {Edmondson}, {Fortson}, {Keel}, {Schawinski}, {Smith}, \&
  {Thomas}}]{masters11}
{Masters}, K.~L., {Nichol}, R.~C., {Hoyle}, B., {et~al.} 2011, \mnras, 411,
  2026

\bibitem[{{Mastropietro} {et~al.}(2005){Mastropietro}, {Moore}, {Mayer},
  {Debattista}, {Piffaretti}, \& {Stadel}}]{mastropietro05}
{Mastropietro}, C., {Moore}, B., {Mayer}, L., {et~al.} 2005, \mnras, 364, 607

\bibitem[{{Mateo}(1998)}]{mateo98}
{Mateo}, M.~L. 1998, \araa, 36, 435

\bibitem[{{Mayer} {et~al.}(2001){Mayer}, {Governato}, {Colpi}, {Moore},
  {Quinn}, {Wadsley}, {Stadel}, \& {Lake}}]{mayer01}
{Mayer}, L., {Governato}, F., {Colpi}, M., {et~al.} 2001, \apj, 559, 754

\bibitem[{{Mayer} {et~al.}(2006){Mayer}, {Mastropietro}, {Wadsley}, {Stadel},
  \& {Moore}}]{mayer06}
{Mayer}, L., {Mastropietro}, C., {Wadsley}, J., {Stadel}, J., \& {Moore}, B.
  2006, \mnras, 369, 1021

\bibitem[{{McConnachie}(2012)}]{mcconnachie12}
{McConnachie}, A.~W. 2012, \aj, 144, 4

\bibitem[{{M{\'e}ndez-Abreu} {et~al.}(2012){M{\'e}ndez-Abreu},
  {S{\'a}nchez-Janssen}, {Aguerri}, {Corsini}, \& {Zarattini}}]{mendez-abreu12}
{M{\'e}ndez-Abreu}, J., {S{\'a}nchez-Janssen}, R., {Aguerri}, J.~A.~L.,
  {Corsini}, E.~M., \& {Zarattini}, S. 2012, \apjl, 761, L6

\bibitem[{{Miller} {et~al.}(1970){Miller}, {Prendergast}, \&
  {Quirk}}]{miller70}
{Miller}, R.~H., {Prendergast}, K.~H., \& {Quirk}, W.~J. 1970, \apj, 161, 903

\bibitem[{{Miller} \& {Smith}(1979)}]{miller_smith79}
{Miller}, R.~H., \& {Smith}, B.~F. 1979, \apj, 227, 785

\bibitem[{{Miwa} \& {Noguchi}(1998)}]{miwa_noguchi98}
{Miwa}, T., \& {Noguchi}, M. 1998, \apj, 499, 149

\bibitem[{{Monaghan}(1992)}]{monaghan92}
{Monaghan}, J.~J. 1992, \araa, 30, 543

\bibitem[{{Mu{\~n}oz} {et~al.}(2010){Mu{\~n}oz}, {Geha}, \&
  {Willman}}]{munoz10}
{Mu{\~n}oz}, R.~R., {Geha}, M., \& {Willman}, B. 2010, \aj, 140, 138

\bibitem[{{Navarro} {et~al.}(1995){Navarro}, {Frenk}, \& {White}}]{nfw95}
{Navarro}, J.~F., {Frenk}, C.~S., \& {White}, S.~D.~M. 1995, \mnras, 275, 720

\bibitem[{{Noguchi}(1987)}]{noguchi87}
{Noguchi}, M. 1987, \mnras, 228, 635

\bibitem[{{Ostriker} \& {Peebles}(1973)}]{ostriker_peebles73}
{Ostriker}, J.~P., \& {Peebles}, P.~J.~E. 1973, \apj, 186, 467

\bibitem[{{Papastergis} {et~al.}(2012){Papastergis}, {Cattaneo}, {Huang},
  {Giovanelli}, \& {Haynes}}]{papastergis12}
{Papastergis}, E., {Cattaneo}, A., {Huang}, S., {Giovanelli}, R., \& {Haynes},
  M.~P. 2012, \apj, 759, 138

\bibitem[{{Pardy} {et~al.}(2016){Pardy}, {D'Onghia}, {Athanassoula}, {Wilcots},
  \& {Sheth}}]{pardy16}
{Pardy}, S.~A., {D'Onghia}, E., {Athanassoula}, E., {Wilcots}, E.~M., \&
  {Sheth}, K. 2016, \apj, 827, 149

\bibitem[{{Patsis} \& {Athanassoula}(2000)}]{patsis_athanassoula00}
{Patsis}, P.~A., \& {Athanassoula}, E. 2000, \aap, 358, 45

\bibitem[{{Peschken} \& {{\L}okas}(2018)}]{peschken_lokas18}
{Peschken}, N., \& {{\L}okas}, E.~L. 2018, submitted to \mnras,
  arXiv:1804.06241

\bibitem[{{Pettitt} \& {Wadsley}(2018)}]{pettitt_wadsley18}
{Pettitt}, A.~R., \& {Wadsley}, J.~W. 2018, \mnras, 474, 5645

\bibitem[{{Pfenniger} \& {Norman}(1990)}]{pfenniger_norman90}
{Pfenniger}, D., \& {Norman}, C. 1990, \apj, 363, 391

\bibitem[{{Power} {et~al.}(2003){Power}, {Navarro}, {Jenkins}, {Frenk},
  {White}, {Springel}, {Stadel}, \& {Quinn}}]{power03}
{Power}, C., {Navarro}, J.~F., {Jenkins}, A., {et~al.} 2003, \mnras, 338, 14

\bibitem[{{Raha} {et~al.}(1991){Raha}, {Sellwood}, {James}, \& {Kahn}}]{raha91}
{Raha}, N., {Sellwood}, J.~A., {James}, R.~A., \& {Kahn}, F.~D. 1991, \nat,
  352, 411

\bibitem[{{Romeo} \& {Wiegert}(2011)}]{romeo_wiegert11}
{Romeo}, A.~B., \& {Wiegert}, J. 2011, \mnras, 416, 1191

\bibitem[{{Saha} \& {Naab}(2013)}]{saha_naab13}
{Saha}, K., \& {Naab}, T. 2013, \mnras, 434, 1287

\bibitem[{{Sales} {et~al.}(2015){Sales}, {Vogelsberger}, {Genel}, {Torrey},
  {Nelson}, {Rodriguez-Gomez}, {Wang}, {Pillepich}, {Sijacki}, {Springel}, \&
  {Hernquist}}]{sales15}
{Sales}, L.~V., {Vogelsberger}, M., {Genel}, S., {et~al.} 2015, \mnras, 447, L6

\bibitem[{{Salo}(1991)}]{salo91}
{Salo}, H. 1991, \aap, 243, 118

\bibitem[{{Sanders} \& {Huntley}(1976)}]{sanders_huntley76}
{Sanders}, R.~H., \& {Huntley}, J.~M. 1976, \apj, 209, 53

\bibitem[{{Scannapieco} \& {Athanassoula}(2012)}]{scannapieco_athanassoula12}
{Scannapieco}, C., \& {Athanassoula}, E. 2012, \mnras, 425, L10

\bibitem[{{Sellwood}(2014)}]{sellwood14}
{Sellwood}, J.~A. 2014, Reviews of Modern Physics, 86, 1

\bibitem[{{Semczuk} {et~al.}(2017){Semczuk}, {{\L}okas}, \& {del
  Pino}}]{semczuk17}
{Semczuk}, M., {{\L}okas}, E.~L., \& {del Pino}, A. 2017, \apj, 834, 7

\bibitem[{{Semczuk} {et~al.}(2018){Semczuk}, {{\L}okas}, {Salomon},
  {Athanassoula}, \& {D{\textquoteright}Onghia}}]{semczuk18}
{Semczuk}, M., {{\L}okas}, E.~L., {Salomon}, J.-B., {Athanassoula}, E., \&
  {D{\textquoteright}Onghia}, E. 2018, \apj, 864, 34

\bibitem[{{Sheth} {et~al.}(2008){Sheth}, {Elmegreen}, {Elmegreen}, {Capak},
  {Abraham}, {Athanassoula}, {Ellis}, {Mobasher}, {Salvato}, {Schinnerer},
  {Scoville}, {Spalsbury}, {Strubbe}, {Carollo}, {Rich}, \& {West}}]{sheth08}
{Sheth}, K., {Elmegreen}, D.~M., {Elmegreen}, B.~G., {et~al.} 2008, \apj, 675,
  1141

\bibitem[{{Shlosman} \& {Noguchi}(1993)}]{shlosman_noguchi93}
{Shlosman}, I., \& {Noguchi}, M. 1993, \apj, 414, 474

\bibitem[{{Simpson} {et~al.}(2018){Simpson}, {Grand}, {G{\'o}mez}, {Marinacci},
  {Pakmor}, {Springel}, {Campbell}, \& {Frenk}}]{simpson18}
{Simpson}, C.~M., {Grand}, R. J.~J., {G{\'o}mez}, F.~A., {et~al.} 2018, \mnras,
  478, 548

\bibitem[{{Skibba} {et~al.}(2012){Skibba}, {Masters}, {Nichol}, {Zehavi},
  {Hoyle}, {Edmondson}, {Bamford}, {Cardamone}, {Keel}, {Lintott}, \&
  {Schawinski}}]{skibba12}
{Skibba}, R.~A., {Masters}, K.~L., {Nichol}, R.~C., {et~al.} 2012, \mnras, 423,
  1485

\bibitem[{{Sormani} {et~al.}(2015){Sormani}, {Binney}, \&
  {Magorrian}}]{sormani15}
{Sormani}, M.~C., {Binney}, J., \& {Magorrian}, J. 2015, \mnras, 449, 2421

\bibitem[{{Spinoso} {et~al.}(2017){Spinoso}, {Bonoli}, {Dotti}, {Mayer},
  {Madau}, \& {Bellovary}}]{spinoso17}
{Spinoso}, D., {Bonoli}, S., {Dotti}, M., {et~al.} 2017, \mnras, 465, 3729

\bibitem[{{Springel}(2005)}]{springel05}
{Springel}, V. 2005, \mnras, 364, 1105

\bibitem[{{Springel} \& {Hernquist}(2003)}]{springel_hernquist03}
{Springel}, V., \& {Hernquist}, L. 2003, \mnras, 339, 289

\bibitem[{{Staveley-Smith} {et~al.}(2003){Staveley-Smith}, {Kim}, {Calabretta},
  {Haynes}, \& {Kesteven}}]{staveley-smith03}
{Staveley-Smith}, L., {Kim}, S., {Calabretta}, M.~R., {Haynes}, R.~F., \&
  {Kesteven}, M.~J. 2003, \mnras, 339, 87

\bibitem[{{Sundin} {et~al.}(1993){Sundin}, {Donner}, \& {Sundelius}}]{sundin93}
{Sundin}, M., {Donner}, K.~J., \& {Sundelius}, B. 1993, \aap, 280, 105

\bibitem[{{Sundin} \& {Sundelius}(1991)}]{sundin_sundelius91}
{Sundin}, M., \& {Sundelius}, B. 1991, \aap, 245, L5

\bibitem[{{Tolstoy} {et~al.}(2009){Tolstoy}, {Hill}, \& {Tosi}}]{tolstoy09}
{Tolstoy}, E., {Hill}, V., \& {Tosi}, M. 2009, \araa, 47, 371

\bibitem[{{Toomre} \& {Toomre}(1972)}]{toomre_toomre72}
{Toomre}, A., \& {Toomre}, J. 1972, \apj, 178, 623

\bibitem[{{Villa-Vargas} {et~al.}(2010){Villa-Vargas}, {Shlosman}, \&
  {Heller}}]{villa_vargas10}
{Villa-Vargas}, J., {Shlosman}, I., \& {Heller}, C. 2010, \apj, 719, 1470

\bibitem[{{Villalobos} {et~al.}(2012){Villalobos}, {De Lucia}, {Borgani}, \&
  {Murante}}]{villalobos12}
{Villalobos}, {\'A}., {De Lucia}, G., {Borgani}, S., \& {Murante}, G. 2012,
  \mnras, 424, 2401

\bibitem[{{Widrow} \& {Dubinski}(2005)}]{widrow_dubinski05}
{Widrow}, L.~M., \& {Dubinski}, J. 2005, \apj, 631, 838

\bibitem[{{Widrow} {et~al.}(2008){Widrow}, {Pym}, \& {Dubinski}}]{widrow08}
{Widrow}, L.~M., {Pym}, B., \& {Dubinski}, J. 2008, \apj, 679, 1239

\bibitem[{{Wilcots}(2009)}]{wilcots09}
{Wilcots}, E.~M. 2009, in IAU Symposium, Vol. 256, The Magellanic System:
  Stars, Gas, and Galaxies, ed. J.~T. {Van Loon} \& J.~M. {Oliveira}, 461--472

\bibitem[{{Zana} {et~al.}(2018){Zana}, {Dotti}, {Capelo}, {Bonoli}, {Haardt},
  {Mayer}, \& {Spinoso}}]{zana18}
{Zana}, T., {Dotti}, M., {Capelo}, P.~R., {et~al.} 2018, \mnras, 473, 2608

\bibitem[{{Zemp} {et~al.}(2011){Zemp}, {Gnedin}, {Gnedin}, \&
  {Kravtsov}}]{zemp11}
{Zemp}, M., {Gnedin}, O.~Y., {Gnedin}, N.~Y., \& {Kravtsov}, A.~V. 2011, \apjs,
  197, 30

\end{thebibliography}

\end{document}